\documentclass[prb,preprint,titlepage,amsmath,amssymb,superscriptaddress,longbibliography]{revtex4-1}

\usepackage{graphicx}
\usepackage{dcolumn}
\usepackage{bm}

\usepackage{soul}

\usepackage[inner=1.5cm,outer=1.5cm,bottom=2cm,tmargin=1cm]{geometry}

\usepackage{mathtools}

\usepackage{outlines}
\usepackage{fancyvrb}
\usepackage{amsmath,nccmath}
\usepackage{hyperref}
\usepackage{braket}
\usepackage{bbold}

\usepackage{amssymb}

\usepackage[usenames, dvipsnames]{color}

\usepackage{float}

\usepackage{rotating}

\usepackage{tensor}

\usepackage{dsfont}
\usepackage{physics}

\usepackage{diagbox}
\usepackage{stackengine}
\newcommand\xrowht[2][0]{\addstackgap[.5\dimexpr#2\relax]{\vphantom{#1}}}

\begin{document}
\title{Subcycle tomography of quantum light}
\author{Geehyun Yang}

\affiliation{Department of Physics, KAIST, Daejeon, Republic of Korea}

\author{Matthias Kizmann}
 \email{mkizmann@uci.edu}
 \affiliation{Department of Chemistry and Physics and Astronomy, University of California, Irvine, USA}

\affiliation{Department of Physics and Center for Applied
Photonics, University of Konstanz,  Germany}
\author{Alfred Leitenstorfer}

\affiliation{Department of Physics and Center for Applied
Photonics, University of Konstanz,  Germany}
\author{Andrey S. Moskalenko}

\email{moskalenko@kaist.ac.kr }

\affiliation{Department of Physics, KAIST, Daejeon, Republic of Korea}

\begin{abstract}

Quantum light is considered to be one of the key resources of the coming second quantum revolution expected to give rise to groundbreaking technologies and applications.
If the spatio-temporal and polarization structure of modes is known, the properties of quantum light are well understood. This information provides the basis for contemporary quantum optics and its applications in quantum communication and metrology.
However, thinking about quantum light at the
most fundamental timescale,
namely the oscillation cycle of a mode or the inverse frequency of an involved photon, we realize that the corresponding picture has been missing until now. For instance, how to comprehend and characterize a single photon at this timescale?
To fill this gap, we demonstrate theoretically how local quantum measurements allow to reconstruct and visualize a quantum field under study at subcycle scales, even when its temporal mode structure is \textit{a priori} unknown. In particular, generation and tomography of ultrabroadband squeezed states as well as photon-subtracted states derived from them are described, incorporating also single-photon states. Our results set a cornerstone in the emerging chapter of quantum physics
termed time-domain quantum optics.
We expect this development to elicit new spectroscopic concepts for approaching e.g.
fundamental correlations and entanglement in the dynamics of quantum matter, overcoming the temporal limitation set by the oscillation cycles of both light and elementary excitations.

\end{abstract}

\maketitle

\section{Introduction}

The quantum nature of light is at the heart of the anticipated advantage of a plethora of emerging quantum technologies. In particular, utilization of so-called squeezed light enables higher sensitivity than possible classically \cite{Caves1981,Xiao1987,Grangier1987,LIGOgravitationalwave}.  Bare vacuum fluctuations are deemed to provide a controlled source of randomness for photonic probabilistic computing \cite{Roques_Carmes2023}.
Squeezed and single-photon states of light are considered as key resources for universal \cite{Menicucci2006,Braunstein2005} and non-universal \cite{Spring2013,Tillmann2013,Crespi2013,Spagnolo2014} quantum computing based on photonics platforms.
More peculiar quantum states of light, such as NOON states, are envisaged to have a potential to be game changers in lithography \cite{Boto2000} and
radar technology \cite{Maccone2020}.
Further, quantum light might be essential for testing of fundamental physical theories such as quantum gravity \cite{Berchera2013}.
Whereas a full description of quantum states of light is provided by the corresponding density matrix, their visualization works in the best way in the phase space of the natural variables of the electromagnetic field by introducing phase-space distributions \cite{Schleich_book}. With respect to other possible phase-space distributions \cite{Sudarshan1963,Glauber1963,Husimi1940} the Wigner function \cite{Wigner1932} stands out by its intuitiveness and transparent connections to observables characterizing quantum-informational resources \cite{Tilma2016}.
Since the density matrix can be reconstructed from the Wigner function  \cite{Tilma2016}, the latter provides a complete description of the field. Despite this fact, till now the concept of the Wigner function in quantum optics has been missing the ultimate temporal resolution,
%
%
which is rooted in its conventional experimental acquisition implying effective averaging over multiple cycles of oscillations of light \cite{Lvovsky2009,Kirchmair2013}.
Here we introduce phase-space distributions of quantum light based on subcycle-resolved tomography. We consider subcycle-squeezed vacuum states of light as well as non-Gaussian states derived from them and demonstrate how their 
evolution
can be tracked in the phase space pushing the temporal resolution to its physical limit.
Our study
provides a cornerstone
for establishing quantum optics in the time domain and characterization of novel quantum resources for unraveling ultrafast quantum phenomena in quantum matter \cite{Masson2022,Boschini2018,Baykusheva2022}.

We want to show how to address key features of quantum states, such as squeezing and negativity of the Wigner function \cite{KenfackNegativity,NegativityQC}, in the subcycle regime, i.e. at timescales below the characteristic optical cycle of the studied light. Currently, sensitive subcycle detection of classical electric fields is possible for signals in the THz or mid-infrared (MIR) frequency ranges. Here, a near-infrared (NIR) probe exhibits a duration shorter than a half-cycle of the signal
(a NIR-signal, NIR/visible-probe version has been also demonstrated recently \cite{Keiber2016}). The full characterization in terms of a (non-monochromatic) classical phasor dynamics in the phase space of the field quadratures is available by means of electro-optic sampling (EOS) \cite{Gallot1999,Sulzer2020}.
The transfer of the advances in ultrabroadband THz technology to the quantum domain
has led to the first measurements of vacuum electric-field fluctuations \cite{Riek2015} and correlations \cite{Benea2019,Settembrini2022} in free space. Moreover, it has been demonstrated theoretically that the dynamics of electric-field variances of ultrashort squeezed vacuum states can be traced directly in the time domain \cite{RiekSubcycle,KizmannSubcycle}. This is in strong contrast to the conventional detection approaches in quantum optics,
where the central frequency of all involved fields is well defined so that the dynamics of the studied quantum field is quasi-stationary on the scale of its characteristic cycle. Thus, broadband quantum light has been mostly investigated by mode matching to a set of separable modes \cite{RoslundWavelength,RaNon-Gaussian}. They
are localized both in the time and frequency domains  but have a relatively small bandwidth with respect to the central frequency, as the studied states themselves. However, without \textit{a priori} knowledge about the state of the incoming light, its separability into a chosen complete set of localized modes cannot be assured so that the information on intermode correlations generally would be lost. Moreover, to achieve subcycle resolution an extreme precision in the engineering of the involved probe modes is required, which is unfeasible in conventional
setups. These facts render the newly introduced subcycle time-domain approach as a breakthrough method in ultrafast quantum optics. A crucial milestone in its formulation shall be the introduction of an appropriate time-dependent phase-space distribution  together with a guideline to its measurement.

Here, we present a new tomography scheme to completely resolve the local
structure of ultrabroadband quantum light in the time domain. The function reconstructed from the tomography protocol represents a time-dependent joint
quasiprobability distribution of the sampled electric field and its conjugate
quadrature \cite{Sulzer2020}, constituting two coordinates of the phase space of the field, in the classical case -- real and imaginary parts of the field phasor. This function is capable to  visualize the dynamics of the field state, providing direct access to the ultrafast evolution of its characteristic features, such as photon content, squeezing and negativity.
To capture the local
structure of ultrafast quantum light,
temporal gating with a subcycle duration with respect to the studied signal is required. This gating can be realized via a nonlinear interaction between the signal and a probe pulse, as in the
quantum extension of EOS \cite{Riek2015,MoskalenkoParaxial,RiekSubcycle,Benea2019,KizmannSubcycle,Settembrini2022,deLiberato2019,Lindel2021}. However, to illuminate the fundamentals of our scheme in the most transparent way, in the main part of this paper we base our considerations on an alternative, linear implementation utilizing a subcycle version of the conventional balanced homodyne detection (BHD) \cite{Vogel2006} technique. In this way the issue of the admixture of the probe shot noise to the detected signal, which is omnipresent in the nonlinear implementation, can be avoided. The dynamics of the probability distributions of all generalized quadratures for the analyzed quantum light is revealed by imposing a controlled time delay and carrier-envelope phase (CEP) shift on the gating local oscillator (LO) pulse, respectively. We show that the collected information enables a reconstruction of the time-resolved Wigner function (TRWF) for all time moments.
%
%
Technical drawbacks might be the difficulty to generate the required ultrashort LO pulses and the current absence of efficient MIR photon detectors but they might be overcome in the near future. The discussion of the nonlinear and experimentally viable version of the scheme, based on EOS, is analogous and provided in Supplementary Information.


\section{Results}
\begin{figure}[t]\includegraphics[width=18cm]{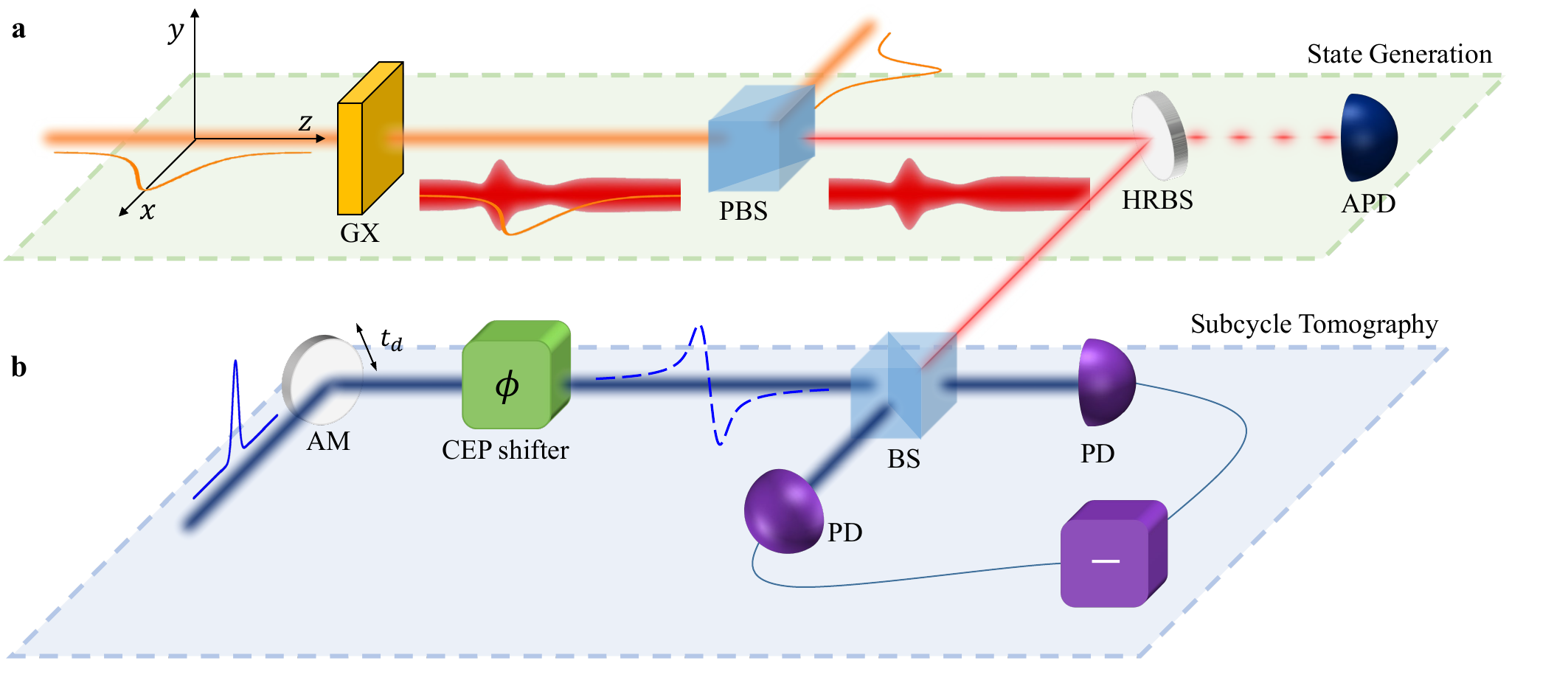}
	\caption{
	\label{Fig:Subcycle Tomography by BHD}
\textbf{Schematic of generation and subcycle tomography of pulsed ultrabroadband non-Gaussian states of quantum light}.
\textbf{a}, Generation of a pulsed squeezed vacuum state with a single-photon subtraction.
A strong coherent mid-infrared (MIR) pulse (orange line) interacts with co-propagating
vacuum fluctuations in a nonlinear generating crystal (GX) to produce an ultrabroadband squeezed MIR vacuum state (red band). The MIR and squeezed vacuum fields have mutually perpendicular polarizations (in the shown example along the $x$- and $y$-directions, respectively) so that the former can be filtered out by a polarizing beam splitter (PBS). The following high-reflectance beam splitter (HRBS) with an avalanche photodiode (APD) is used to subtract a single photon.
\textbf{b}, Subcycle tomography. The
temporal dynamics of the generated state is captured by homodyne detection employing a half-cycle local oscillator (LO) pulse (blue line) whose relative time delay $t_d$  and carrier-envelope phase (CEP) $\phi$ can be modulated by an adjustable mirror (AM) and a CEP shifter, respectively. Blue dashed line after the CEP shifter depicts the LO with $\phi=-\frac{\pi}{2}$. The outcome resulting from the superposition of the LO with the generated pulsed quantum light at a 50:50 beam splitter (BS) is analyzed by balanced photodetectors (PDs).}
\end{figure}

The schematic of the proposed setup for the generation and measurement of ultrashort quantum states is illustrated in Fig.~\ref{Fig:Subcycle Tomography by BHD}. Its upper part, Fig.~\ref{Fig:Subcycle Tomography by BHD}a, corresponds to the generation process. Here
a strong linearly polarized coherent pulse $\mathcal{E}(t)$ (orange) interacts with the co-propagating vacuum field through a $\chi^{(2)}$ nonlinear process in the generating crystal (GX), giving rise to an ultrashort pulse of the squeezed vacuum light with electrical field operator  $\hat{E}_\mathrm{psq}(t)$ (red) at the perpendicular polarization \cite{KizmannSubcycle}. Considering a thin GX, which in terms of   phase matching \cite{Boyd_book} can be treated as effectively dispersionless, $\chi^{(2)}$ is taken as a constant. For our illustrative calculations, we assume that $\mathcal{E}(t)$ corresponds to a MIR half-cycle pulse (HCP) \cite{Arkhipov_HCP_Review,Moskalenko2017} and GX consists of a ZnTe crystal, which is thinner than $15\,\mu$m\cite{KizmannSubcycle}. The generated quantum field $\hat{E}_\mathrm{psq}(t)$ is separated from the coherent drive $\mathcal{E}(t)$ by passing through a polarizing beam splitter (PBS). An optional postselection of the signal based on the positive response of an avalanche photodiode (APD) placed behind a high-reflectance beam splitter (HRBS), enables subtraction of a single photon from $\hat{E}_\mathrm{psq}(t)$\cite{Averchenko2016}. Mode-independent high reflectance of the HRBS and perfect quantum efficiency of the APD are assumed to avoid temporal distortions upon postselection. Realizing this option
imposes  non-Gaussianity on the output pulsed quantum light.

The detection part implementing the subcycle tomography based on the pulsed BHD is illustrated in  Fig.~\ref{Fig:Subcycle Tomography by BHD}b.
To obtain time-resolved information on the generated pulsed quantum light $\hat{E}(t)$, a strong unipolar LO
$R(t)$ shorter than the driving field $\mathcal{E}(t)$ is used. The time delay $t_d$ between the LO and $\hat{E}(t)$ can be controlled by an adjustable mirror (AM), whereas the CEP shifter implements the corresponding phase change $\phi$ on the LO.
Varying
$t_d$
allows for the sampling of the quantum electric field $\hat{E}(t)$ at different times, while the CEP shifter plays a similar role as the phase shifter in conventional
BHD.
In result, we get access to the quantum statistics of the signal $\hat{E}^{(d)}_{\phi}(t_d)=\int_{-\infty}^{\infty} \mathrm{d}t\,R_{-\phi}(t-t_d)\hat{E}(t)$, where $R_\phi(t)$ is the gating function CEP shifted by $\phi$. This signal can be rewritten as $\hat{E}^{(d)}_{\phi}(t_d)=\int_{-\infty}^{\infty} \mathrm{d}t\,R(t-t_d)\hat{E}_{\phi}(t)$,
corresponding to the locally detected CEP-shifted electric field.
The blue dashed wave form in Fig.~\ref{Fig:Subcycle Tomography by BHD}b represents $R_{-\frac{\pi}{2}}(t)$,
i.e. the LO transient for the detection of the $\frac{\pi}{2}$-CEP-shifted electric field. Note that the same information about the quantum statistics of $\hat{E}(t)$ can also be obtained using electro-optic sampling\cite{Riek2015,RiekSubcycle,Benea2019}. A detailed description of this measurement
option
is provided in Supplementary Information.
Normalizing $\hat{E}^{(d)}(t_d)$ and $\hat{E}^{(d)}_{\frac{\pi}{2}}(t_d)$ to satisfy the canonical commutation relation, 
we introduce time-resolved
conjugate quadratures
\begin{eqnarray}\label{eq:quadratures}
\hat{X}(t_d)=-
\mathcal{N}\hat{E}^{(d)}_{\frac{\pi}{2}}(t_d),\qquad \hat{P}(t_d)=-
\mathcal{N}
\hat{E}^{(d)}(t_d),
\end{eqnarray}
where the normalization coefficient
$\mathcal{N}=1/\sqrt{\big\lvert \big[\hat{E}^{(d)}(t_d),\hat{E}^{(d)}_{\frac{\pi}{2}}(t_d)\big]\big\rvert}$
does not depend on the time delay (see Methods) and $[\hat{X}(t_d),\hat{P}(t_d)]=1$. Like for the usual frequency-domain Wigner function, for any state of light, determined by its density matrix $\hat{\rho}$, we can introduce the corresponding characteristic function
\begin{eqnarray}\label{eq:characteristicfunction}
		\widetilde{W}(u,v;t_d)\coloneqq\text{tr}\Big[\hat{\rho}\;\text{exp}\big\{-iu\hat{X}(t_d)-iv\hat{P}(t_d)\big\}\Big].
\end{eqnarray}
Its Fourier transform $W(x,p;t_d)=(2\pi)^{-2}\int_{-\infty}^{\infty}\int_{-\infty}^{\infty}\mathrm{d}u\mathrm{d}v\,\widetilde{W}(u,v;t_d)e^{iux+ivp}$ we term 
time-resolved Wigner function (TRWF).
The TRWF represents a quasiprobability distribution.
It captures the complete information about the sampled electromagnetic field, which is contained in the positive-frequency part of the electric-field operator $\hat{E}^{(+)}(t)$, with a temporal resolution corresponding to the duration of the probe pulse (see Methods).
Since the center of gravity of the TRWF along the $p$-axis ($x$-axis) determines the detected real (imaginary) value of the complex classical electric field $E^{(+)}(t)$ at $t=t_d$,
the TRWF at each given time may be regarded as a generalization of the classical field phasor.

\begin{figure}[t]
\includegraphics[width=18cm]{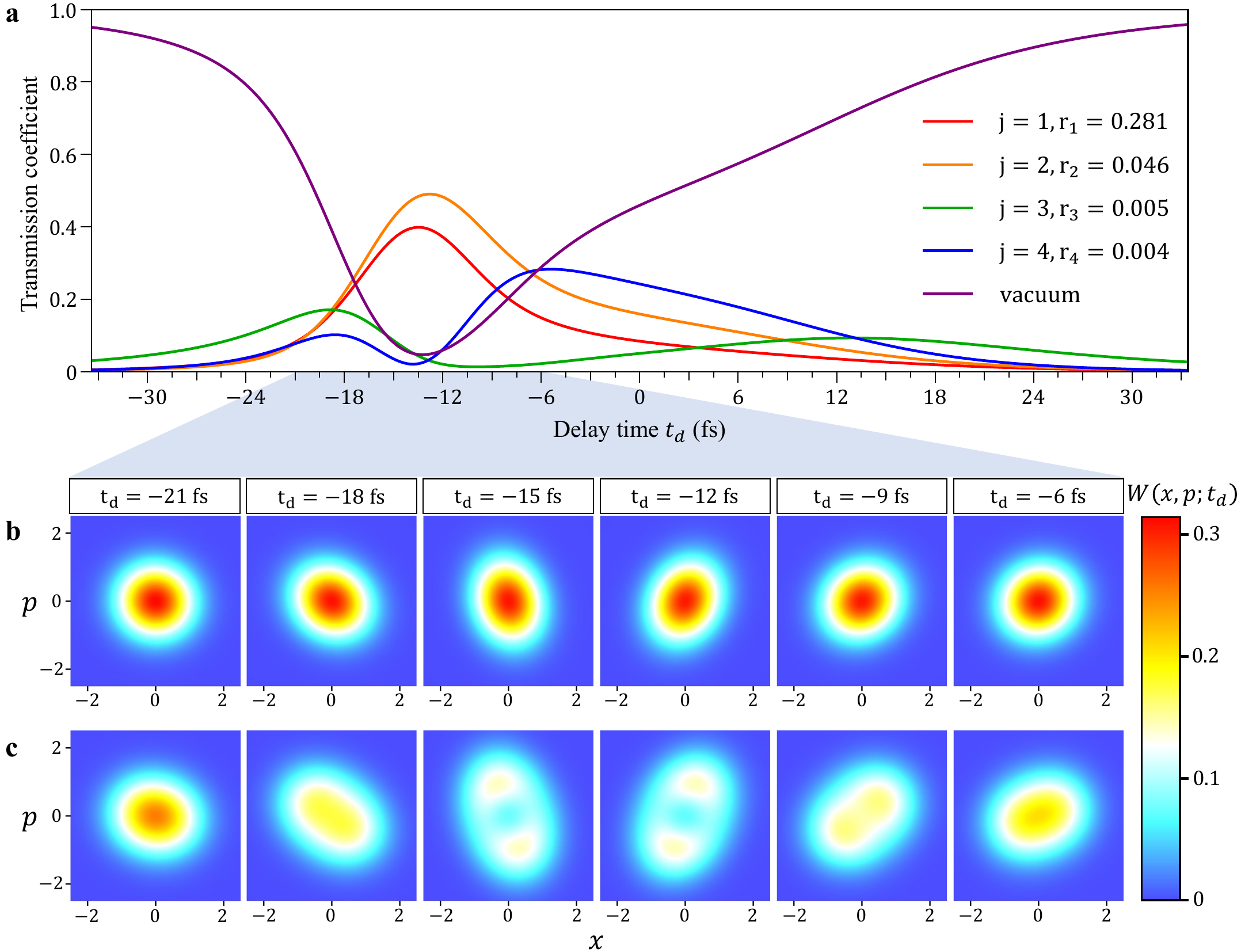}
	\centering
	\caption{\label{Fig:TRWF_dynamics}
		\textbf{Subcycle dynamics of the pulsed squeezed state and single-photon-subtracted state produced from it.} \textbf{a,} Transmission to the detection mode from four dominant separable modes and the remaining part (regarded as vacuum) of the pulsed squeezed state, shown in dependence on the time delay $t_d$.  Each mode is represented by a squeezing parameter $r$, where the vacuum means $r=0$. \textbf{b,c,}
  Snapshots of the time-resolved Wigner function (TRWF) are shown for several values of $t_d$ for the pulsed squeezed state (\textbf{b}) and the corresponding photon-subtracted state (\textbf{c}). The horizontal (vertical) axis represents the $\hat{x}$ ($\hat{p}$) quadrature which is proportional to $-\hat{E}^{(d)}_{\frac{\pi}{2}}$ ($-\hat{E}^{(d)}$). Parameters: $r_\mathrm{eff}=5$, $\delta_d=16$~fs, and $\delta_p=5.8$~fs (cf. Methods).}
\end{figure}

To illustrate the evolution of the TRWF, we first
consider a pulsed squeezed vacuum state produced in the upper branch of the setup shown in Fig.~\ref{Fig:Subcycle Tomography by BHD}. In the frequency domain, the corresponding field is generally given by the continuous Bogoliubov transformation:
$\hat{b}(\omega)=\int_0^\infty\mathrm{d}\omega'\, p(\omega,\omega')\hat{a}(\omega')+\int_0^\infty\mathrm{d}\omega'\, q(\omega,\omega')\hat{a}^\dagger(\omega')$,
where $\hat{a}(\omega)$ and $\hat{b}(\omega)$ are the photon annihilation operators of the input and output field, respectively. The particular forms of
$p(\omega,\omega')$ and $q(\omega,\omega')$ are determined by the shape of the driving $\mathcal{E}(t)$ (see Methods).
The calculated TRWF $W_\mathrm{psq}(x,p;t_d)$ resulting from the sampling of such a state is analyzed in Fig.~\ref{Fig:TRWF_dynamics}. In Fig.~\ref{Fig:TRWF_dynamics}b, we see that the distribution, which initially corresponds to a bare vacuum state, gets squeezed along a particular axis in the phase space. Both the orientation of this axis and the degree of squeezing vary on a femtosecond timescale. After reaching its maximum, the degree of squeezing then starts to decrease, whereas the axis continues to rotate clockwise until the symmetric vacuum distribution is restored.
%
%
To comprehend the
dynamics in detail, we find that it is useful to represent the pulsed squeezed vacuum
as a separable state consisting of a finite number of single-mode squeezed states $\hat{\rho}_\mathrm{psq}=\bigotimes_{j=1}^{n}\hat{\rho}_j$
with squeezing parameters $r_j$ characterizing each single mode.
This decomposition is performed by
the Bloch-Messiah reduction,
leading to the input [output] localized mode basis
$\hat{a}_j=\int_0^\infty\mathrm{d}\omega\,\phi_j(\omega)\hat{a}(\omega)$
$[\hat{b}_j=\int_0^\infty\mathrm{d}\omega\,\psi_j(\omega)\hat{b}(\omega)]$, satisfying $p(\omega,\omega')=\sum_j \psi_j^*(\omega)\cosh r_j\, \phi_j(\omega')$ and $q(\omega,\omega')=\sum_j \psi_j^*(\omega)\sinh r_j\,\phi_j^*(\omega')$\cite{McCutcheon2018,Wasilewski2006}. In terms of these modes, the Bogoliubov transformation becomes $\hat{b}_j=\cosh r_j\,\hat{a}_j+\sinh r_j\,\hat{a}_j^\dagger\ $ ($j=1,\ldots,n)$ so that the pulsed squeezed light can be described as a combination of  single-mode squeezed states, corresponding to temporally localized modes\cite{Raymer2020,Fabre2020,Sharapova2018}. By regarding the contributions of all modes with squeezing parameters $r_j$ less than $10^{-3}$ as a vacuum, the description of the squeezing process can be effectively limited to
a small number of principal modes. For example, for the pulsed driving  described in Methods, it is sufficient to consider just four such modes:
\begin{eqnarray}\label{eq:BMreduction}
\hat{\rho}_\mathrm{psq}=\bigotimes_{j=1}^{4}\hat{\rho}_j\otimes \hat{\rho}_\mathrm{vac}\,.
\end{eqnarray}
In this case, the squeezing parameters are $r_1=0.281,\, r_2=0.046,\, r_3=0.005,$ and $r_4=0.004$. Thus, the observed time-dependent squeezing mainly results from the contribution of the first localized principal mode, with the rest of the effect mostly coming from the second mode.

Figure \ref{Fig:TRWF_dynamics}a illustrates how each of the principal modes contributes to the dynamics of the TRWF. To understand these contributions, we note that they are determined by projections of  the detection mode $\hat{A}^{(d)}(t_d)=\frac{\hat{X}(t_d)+i\hat{P}(t_d)}{\sqrt{2}}$, incorporated by the gating, to the principal modes. Specifically, they are quantified by transmission coefficients $T_j(t_d)=\lvert \theta_j(t_d)\rvert^2$ (shown in Fig.~\ref{Fig:TRWF_dynamics}a),
where the commutators $\theta_j(t_d)=[\hat{A}^{(d)}(t_d),\hat{b}^\dagger_j]$ are the projection coefficients \cite{Unruh-DeWitt} of equation \eqref{eq:modaldecompositionofA} in Methods.
Taking into account that $\hat{A}^{(d)}(t_d)$ can be expressed via $\hat{E}^{(+)}(t)$ as $\hat{A}^{(d)}(t_d)=-i\sqrt{2}\mathcal{N}
%
%
\int_{-\infty}^\infty \mathrm{d}t \, R(t-t_d)\hat{E}^{(+)}(t)$,
%
we can write the coefficients $\theta_j(t_d)$ also as
\begin{eqnarray}\label{eq:transmittance}
		\theta_j(t_d)=
-i\sqrt{2}\mathcal{N}
  \int_{-\infty}^\infty \mathrm{d}t \, R(t-t_d)\alpha_j(t),
\end{eqnarray}
where $\alpha_j(t)=[\hat{E}^{(+)}(t),\hat{b}^\dagger_j]\,$ are independent from the particular shape of the probe.
%
%
Hence, contributions of each mode $j$ to the TRWF can be inferred from the temporal shape of the field mode $\alpha_j(t)$, uniquely determined by the generated short-living state, and gating $R(t)$, determined solely by the probe.
%
In the case of a Gaussian probe pulse (explicit form is given in Methods) with duration, $\delta_p=5.8\,$fs shown in Fig.~\ref{Fig:TRWF_dynamics}a, the corresponding time resolution is sufficient to capture $\alpha_1(t)$ and $\alpha_2(t)$ while the features of $\alpha_3(t)$ and $\alpha_4(t)$ are washed out (see Supplementary Information). However, the latter fact does not represent a real problem since most of the generated photons belong to the first and second principal modes.
Equation \eqref{eq:transmittance} implies then $T_j(t_d)=\lvert \theta_j(t_d) \rvert ^2\sim \lvert \alpha_j(t_d) \rvert^2$ for $j=1,2$, meaning that these transmission coefficients closely follow the dynamics of the corresponding field modes.
%
We find
that in the studied case the temporal asymmetry of these coefficients
can be related to the modified effective time flow $\tau(t)$ for the output fields induced by the driving inside the GX, where $\tau(t)$ is termed conformal time \cite{KizmannSubcycle}. That means
the temporal shape of both
$\lvert \alpha_1(t)\rvert$ and $\lvert \alpha_2(t)\rvert$ closely resembles $\mathcal{E}(\tau(t))$ (see Supplementary Information). Especially, near $t_d=-14\,$fs, when $\mathcal{E}(\tau(t))$ becomes significant, high squeezing
 is
observed for the TRWF, resulting from large contributions of the first and second modes.
Negligible contributions emerge from
the third and fourth modes due to the washing out of their
rapid oscillations.
According to equation \eqref{eq:modaldecompositionofA} in Methods, the quadratures of the TRWF can be expanded into the quadratures of the principal modes:
$\left[\begin{array}{cc}\hat{X}(t_d) & \hat{P}(t_d)\end{array}\right]^\top=\sum\limits_{j} \mathbf{O}_j(t_d)\,\left[\begin{array}{cc}\hat{x}_j & \hat{p}_j\end{array}\right]^\top$,
where $\mathbf{O}_j(t_d)=\left[\begin{array}{cc}\Re\theta_j(t_d) & -\Im\theta_j(t_d)\\ \Im\theta_j(t_d) & \Re\theta_j(t_d)\end{array}\right]$
and $^\top$ means transposition. Therefore, the characteristic function defined by equation \eqref{eq:characteristicfunction} can also be decomposed into the principal modes as
\begin{eqnarray}\label{eq:characteristicfunctionofPSQ}
    \widetilde{W}_\mathrm{psq}(u,v;t_d)=\widetilde{W}_\mathrm{vac}\big(\theta_\mathrm{vac}(t_d)\,\left[\begin{array}{cc} u & v\end{array}\right]^\top\big)\prod_{j=1}^4\widetilde{W}_{\mathrm{sq},j}\big(\mathbf{O}^\top_j(t_d)\left[\begin{array}{cc} u & v\end{array}\right]^\top\big),
\end{eqnarray}
where $\theta_\mathrm{vac}(t_d)=\sqrt{1-\sum\limits_{j=1}^4\lvert \theta_j(t_d) \rvert^2}$. $\widetilde{W}_\mathrm{vac}(u,v)$ and $\widetilde{W}_{\mathrm{sq},j}(u,v)$ are the characteristic functions of the vacuum and the squeezed state with squeezing parameter $r_j$, respectively.
For brevity, we have used the convention $\widetilde{W}([u\ v]^\top)\equiv\widetilde{W}(u,v)$.
The Fourier transform of $\widetilde{W}_\mathrm{psq}(u,v;t_d)$ gives then a two-dimensional normal distribution $W_\mathrm{psq}(x,p;t_d)=\frac{1}{2\pi\sqrt{\det \boldsymbol{\Sigma}(t_d)}}\exp\big(-\frac{1}{2}\left[\begin{array}{cc}x & p\end{array}\right]\boldsymbol{\Sigma}^{-1}(t_d)\left[\begin{array}{cc}x & p\end{array}\right]^\top\big)$ with the covariance matrix
\begin{eqnarray}\label{eq:Covariancematrix}
\boldsymbol{\Sigma}(t_d)=\frac{1}{2}\left(\theta^2_\mathrm{vac}(t_d)\,\mathbf{I}+\sum\limits_{j=1}^4\mathbf{O}_j(t_d) \,\mathbf{S}_j \,\mathbf{O}^\top_j(t_d)\right),
\end{eqnarray}
 where $\mathbf{I}=\left[\begin{array}{cc}1 & 0 \\ 0 & 1\end{array}\right]$ and $\mathbf{S}_j=\left[\begin{array}{cc}e^{2r_j} & 0 \\ 0 & e^{-2r_j}\end{array}\right]$. Hence, we see that $W_\mathrm{psq}(x,p;t_d)$ can be computed just from the covariance matrix. Equation \eqref{eq:Covariancematrix} explicitly implies that the evolution of the TRWF $W_\mathrm{psq}(x,p;t_d)$ is mainly affected by the principal modes $j$ with the largest $r_j$ entering the matrix $\mathbf{S}_j$ as mentioned before. In particular, the
 maximal quadrature variance
 and the angle of the corresponding axis reflected in $W_\mathrm{psq}(x,p;t_d)$ can be well approximated by
 $\frac{1}{2}\left[1+T_1(t_d)(e^{2r_1}-1)\right]$
and $\mathrm{arg}(\theta_1(t_d))$, respectively. Here $\mathrm{arg}(z)$ denotes the argument of a complex number $z$. The quadrature variance for the perpendicular axis is given by $\frac{1}{2}\left[1-T_1(t_d)(1-e^{-2r_1})\right]$. Notice that the product of these
variances slightly exceeds $1/4$, as we would have for minimum uncertainty states such as the vacuum. This effect can be mainly attributed to the fact that a certain time-dependent portion of the photons transmitted to the detection mode $\hat{A}^{(d)}(t_d)$ are thermal photons. A related phenomenon was reported for a rapidly switched Unruh-DeWitt detector \cite{Unruh-DeWitt}.
The analysis of the TRWF dynamics considering more modes is provided in Supplementary Information. 

Now we want to consider an example of non-Gaussian pulsed quantum light.
It is described in Fig.~\ref{Fig:Subcycle Tomography by BHD} selecting the option where the APD is used to implement a postselection protocol, which leads to a photon subtraction from the pulsed squeezed state. A signal in the APD originates from annihilation of a photon in one of the principal modes.
The resulting photon-subtracted state is given by (see Supplementary Information) $\hat{\rho}_\mathrm{sub}=\frac{1}{N}\sum\limits_{j=1}^4\hat{b}_j\hat{\rho}_\mathrm{psq}\hat{b}_j^\dagger$, with $N=\sum\limits_{j=1}^4\sinh^2 r_j$ being the number of photons contained in the original squeezed state.
Equation \eqref{eq:characteristicfunction} for $\hat{\rho}_\mathrm{sub}$ leads then to
\begin{equation}\label{eq:characterizticfunctionofsubtratedstate}
\widetilde{W}_\mathrm{sub}(u,v;t_d)=\frac{1}{N}\widetilde{W}_\mathrm{vac}\big(\theta_\mathrm{vac}(t_d) \left[\begin{array}{cc}u & v\end{array}\right]^\top\!\big)
\sum\limits_{\sigma\in P_\mathrm{cyc}}\!\!\widetilde{W}_{\mathrm{sub},\sigma(1)}\big(\mathbf{O}_{\sigma(1)}^\top(t_d) \left[\begin{array}{cc}u & v\end{array}\right]^\top\!\big)\prod\limits_{j=2}^4\widetilde{W}_{\mathrm{sq},\sigma(j)}\big(\mathbf{O}_{\sigma(j)}^\top(t_d) \left[\begin{array}{cc}u & v\end{array}\right]^\top\!\big),
\end{equation}
where $\widetilde{W}_{\mathrm{sub},j}(u,v)$ is the characteristic function of the
corresponding single-photon-subtracted squeezed state $j$
[i.e., $\widetilde{W}(u,v)$ of $\hat{b}_j\rho_j\hat{b}_j^\dagger$\cite{Averchenko2016}] and $P_\mathrm{cyc}$ represents
the set of all possible cyclic permutations $\sigma$ of $\{1,2,3,4\}$.
$W_\mathrm{sub}(x,p;t_d)$ is computed now by Fourier transform of equation \eqref{eq:characterizticfunctionofsubtratedstate}. The postselection eliminates the vacuum in the corresponding original single-mode squeezed state $j$, not affecting the states of the other principal modes, and enforces non-Gaussian features in the TRWF. In the detection mode $\hat{A}^{(d)}(t_d)$, it decreases the contribution of the vacuum state while the admixture of the single-photon state increases (see Supplementary Information). Such a weight shift from the vacuum to the single-photon state appears in the TRWF as a broadening and reduction of its value at the origin (see Fig.~\ref{Fig:TRWF_dynamics}c).
With a uniform transmittance of the HRBS and a relatively higher squeezing parameter of the first mode,
we can ensure that the postselection does not change the dynamics of the squeezing axis and strength. Interestingly, considering the weak squeezing limit realized via a correspondingly small driving field in the same experimental scheme, analogously to the common single-mode case \cite{LeonhardtEssential}, a
single photon
confined to an ultrashort time interval
can be generated. Its dynamics and characteristic features are well captured through the TRWF (see Supplementary Information).

\begin{figure}[t]\includegraphics[width=18cm]{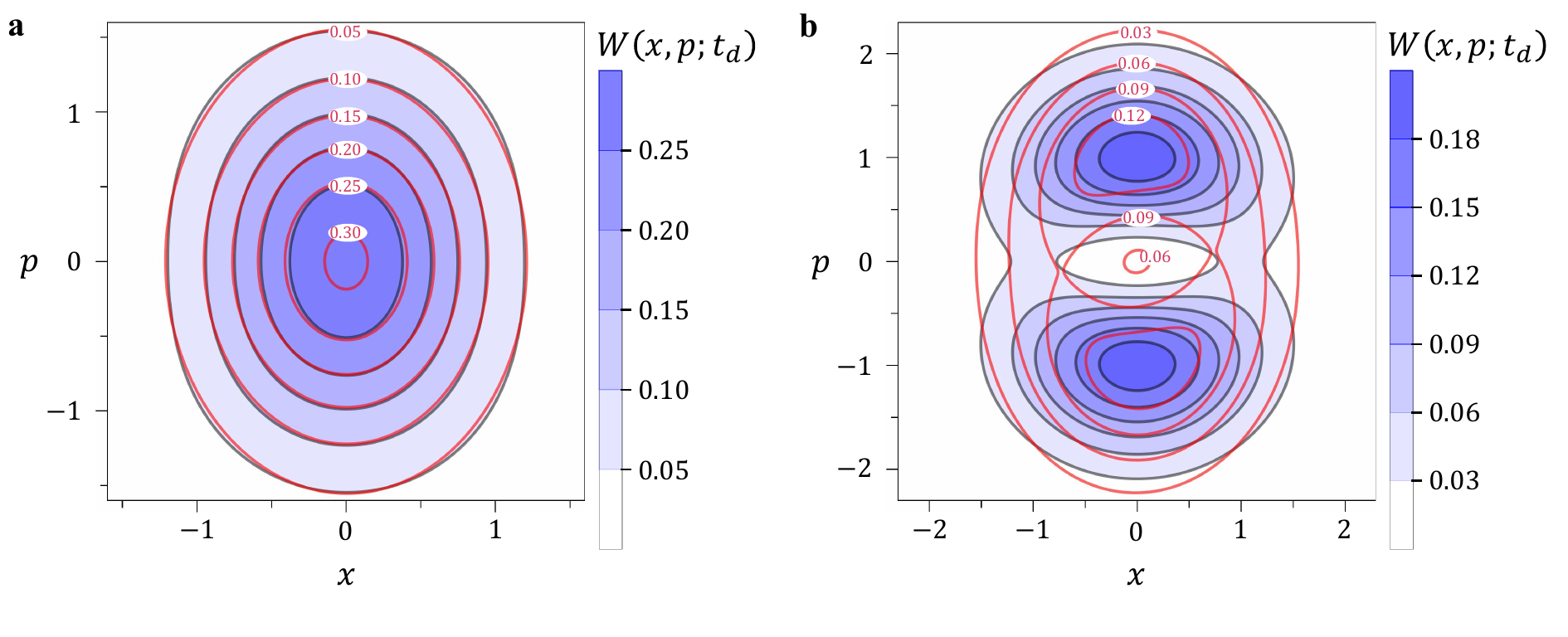}
	\centering
	\caption{\label{Fig:GCexpansion}
		\textbf{Reconstruction of the TRWF from its moments up to the 2nd order (including cross moments).} The red contours without shading correspond to the exactly calculated TRWF whereas black contours with blue shading indicate the result of the reconstruction based on the 2nd-order Gram-Charlier expansion for \textbf{(a)} the pulsed squeezed state at $t_d=-14.1$~fs and \textbf{(b)} the single-photon-subtracted state at $t_d=-13.8$~fs. The chosen values of $t_d$ correspond to the least favorable time moments, i.e.  when the Hilbert-Schmidt distance between the exact solution and its reconstruction is maximal; $D_\mathrm{HS}=0.0007$ for $W_\mathrm{psq}(x,p;t_d)$ and $D_\mathrm{HS}=0.0551$ for $W_\mathrm{sub}(x,p;t_d)$. The parameter values are as in Fig.~\ref{Fig:TRWF_dynamics}.}
\end{figure}

A reconstruction of the TRWF
exploiting
the inverse Radon transform
(see Supplementary Information)
would typically demand a considerable number of measurements
to obtain the full quantum statistical distribution of the quadrature. Moreover, they have to be repeated for
a set of various CEP shifts, which are required to perform the corresponding numerical integration. Then the procedure has to be  repeated also for a large number of time delays assuring the temporal resolution.
An alternative approach to
obtain
the TRWF efficiently can be based on the Gram-Charlier expansion
\eqref{eq:GCexpansionofW} of Methods. To approximate the TRWF to the $N$th order of that expansion, one needs to measure $N+1$ phase-rotated quadratures. Here we limit our consideration to the second-order approximation which already can lead to a very good precision.
To show that we take the introduced TRWF examples for the Gaussian and non-Gaussian cases, $W_\mathrm{psq}(x,p;t_d)$ and $W_\mathrm{sub}(x,p;t_d)$, and perform the estimation from the second moments, i.e., $V_X(t_d)=\langle\hat{X}^2(t_d)\rangle, V_P(t_d)=\langle\hat{P}^2(t_d)\rangle$, and $\mathrm{Cov}_{X,P}(t_d)=\langle \big\{\hat{X}(t_d),\hat{P}(t_d)\big\}/2 \rangle$.
Here $\{,\}$ denotes the anticommutator and we took into account $\langle\hat{X}(t_d)\rangle=\langle\hat{P}(t_d)\rangle=0$. These moments can be obtained by measurements of only three phase-rotated quadratures (see Methods) and lead to
\begin{equation}
W(x,p;t_d)\approx \Big\{\big[2-
V_X(t_d)-V_P(t_d)\big]+4\mathrm{Cov}_{X,P}(t_d)xp
+\big[2V_X(t_d)-1\big]x^2
+\big[2V_P(t_d)-1\big]p^2\Big\}W_\mathrm{vac}(x,p),
\end{equation}
where $W_\mathrm{vac}(x,p)$ refers to
the vacuum. The similarity between any two phase-space distributions can be quantified by the Hilbert-Schmidt distance $D_\mathrm{HS}(W_1,W_2)=2\pi\iint \big[W_1(x,p)-W_2(x,p)\big]^2\mathrm{d}x\mathrm{d}p$\cite{Dodonov2000}. To demonstrate the reliability of the estimate,
the comparison between the exact and estimated $W(x,p;t_d)$ is provided in Fig.~\ref{Fig:GCexpansion} for both the pulsed squeezed state (a) and the  single-photon-subtracted state (b) at the time moment when
$D_\mathrm{HS}$ reaches its maximum. As can be seen from Fig.~\ref{Fig:GCexpansion}a, the estimation works very well for the pulsed squeezed state because $W_\mathrm{psq}(x,p;t_d)$ is quite similar to $W_\mathrm{vac}(x,p)$. This changes only for much higher squeezing strengths requiring then the measurement and inclusion of the higher moments (cf. Supplementary Information).
Although the result of the estimate
for $W_\mathrm{sub}(x,p;t_d)$ in Fig.~\ref{Fig:GCexpansion}b is not as good as the case of the pulsed squeezed state, it still captures the non-Gaussianity and the orientation of the distribution.
Deviations in the estimate
of the TRWF become smaller for other times,
 as
$D_\mathrm{HS}$ decreases.
In addition, the results can be further improved through the inclusion of the higher moments
for the corresponding larger number of the phase-rotated quadratures. Thus, this
method of reconstruction of the TRWF is
a powerful tool
assuring a desired precision by a minimized number of measurements.

\begin{figure}[t]\includegraphics[width=18cm]{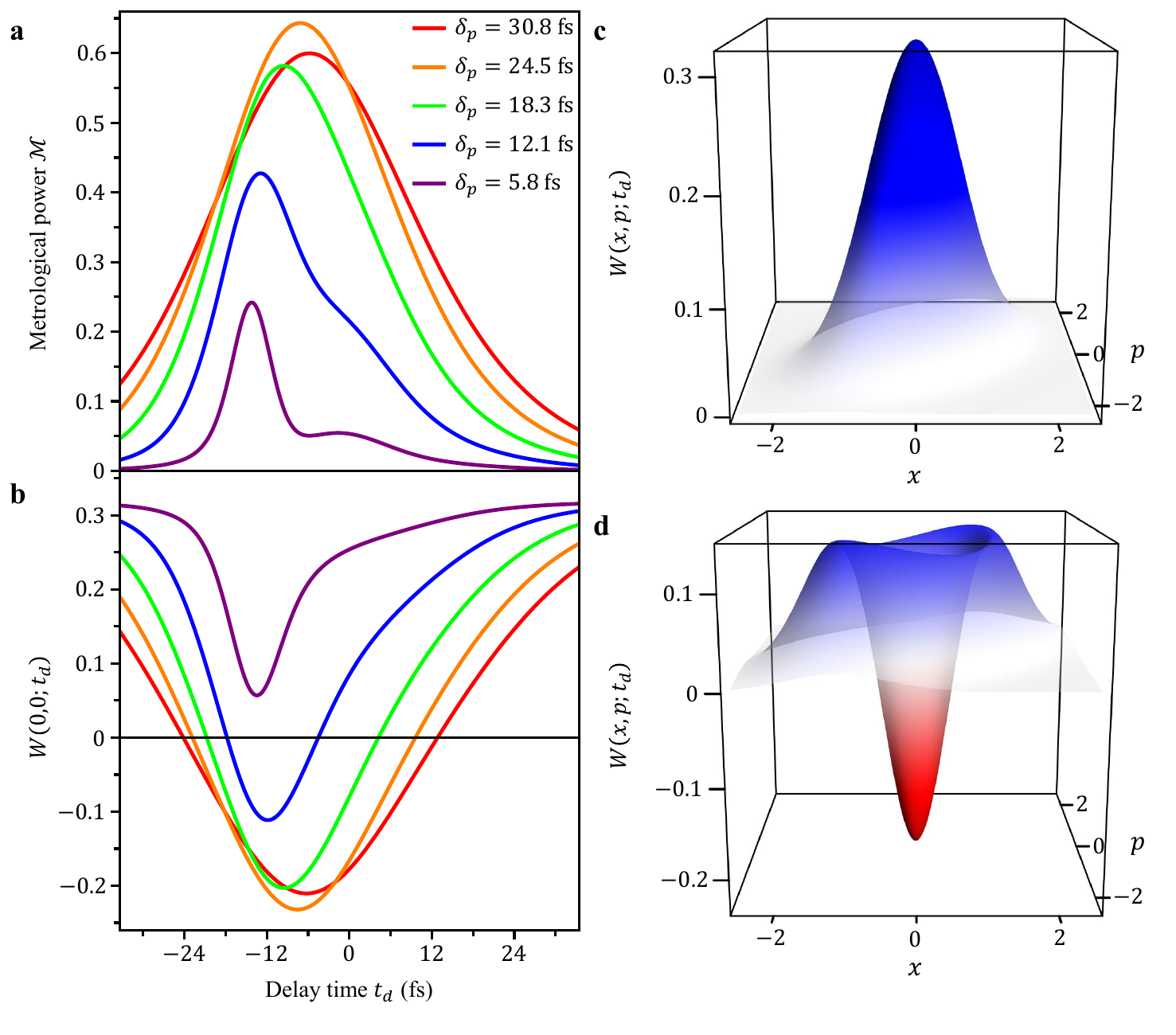}
	\centering
	\caption{\label{Fig:MPandNegativity}
		\textbf{Metrological power and TRWF negativity in dependence on the probe pulse duration.} \textbf{a,} Dynamics of the metrological power $\mathcal{M}$  for the pulsed squeezed state and various values of the probe pulse duration $\delta_p$.
\textbf{b,} Dynamics of the TRWF at the origin $W(0,0;t_d)$ for the single-photon-subtracted state and same values of $\delta_p$.
\textbf{c,} TRWF of the squeezed state at $t_d=-7.1$~fs and $\delta_p=24.5$~fs giving the largest metrological power $\mathcal{M}=0.643$. \textbf{d,} TRWF of the single-photon-subtracted state at $t_d=-7.5$~fs and $\delta_p=24.5$~fs giving the largest negativity of the TRWF at the origin $W_\mathrm{sub}(0,0;t_d)=-0.232$.}
\end{figure}

As fluctuations of the electric field highly affect sensitivity in Michelson interferometers, a variance smaller than that of the vacuum is considered to be necessary to achieve quantum advantage in related metrology schemes \cite{LIGOgravitationalwave}.
Since any positive Wigner function can be emulated classically, another important quantum-informational characteristic is the negativity that should appear to get quantum advantage in quantum computing \cite{NegativityQC}. To
analyze the evolution of these properties with a subcycle resolution, we simulate
the dynamics of the metrological power \cite{MetrologicalPower} $\mathcal{M}=\frac{1}{2}\big(\frac{1}{V_\mathrm{min}}-\frac{1}{V_\mathrm{vac}}\big)$ ($V_\mathrm{min}$ is the minimum variance of the phase-rotated quadratures and $V_\mathrm{vac}$ is its vacuum level) and the TRWF value at the origin $W(0,0;t_d)$ for our
squeezed state (Fig.~\ref{Fig:MPandNegativity}a) and
photon-subtracted state (Fig.~\ref{Fig:MPandNegativity}b), respectively. The temporal resolution is controlled by the probe pulse duration $\delta_p$.
As the squeezing parameters for the first and second principal modes are much larger than for the other modes, $V_\mathrm{min}$ can be approximated from the lower eigenvalue of the covariance matrix $\Sigma(t_d)$ as
\begin{equation}\label{eq:Vminimum}
V_\mathrm{min}\approx\,\frac{1}{2}+T_1(t_d)\,r_1^2-\sqrt{[T_1(t_d)\,r_1+T_2(t_d)\,r_2]^2+2T_1(t_d)T_2(t_d)\,r_1r_2[\cos\varphi(t_d)-1]}\;,
\end{equation}
where $\varphi(t_d)=2[\mathrm{arg}(\theta_1(t_d))-\mathrm{arg}(\theta_2(t_d))]$.
Hence, the metrological power nearly follows the transmission coefficient of the first mode $T_1(t_d)$.  Minor contributions come from the second mode. Especially around $t_d=-8.4\,$fs, where a shallow valley appears in Fig.~\ref{Fig:MPandNegativity}a for $\delta_p=5.8\,$fs, the squeezing character of the contributions to the TRWF from the first and second modes occurs to be opposite with $\mathrm{arg}(\theta_1)-\mathrm{arg}(\theta_2)=\frac{\pi}{2}$, i.e. whereas the first mode is squeezed along the axis corresponding to $V_\mathrm{min}$ the second mode is anti-squeezed. The superposition of these modes causes destructive interference in the resulting squeezing of $W_\mathrm{psq}(x,p;t_d)$ (for more details, see Fig.~\ref{Fig:Dynamics of the squeezed state}b of Supplementary Information). We can see that such destructive interference becomes negligible for large probe pulse durations due to a small transmission of the second mode $T_2(t_d)$. On the other hand, the photon subtraction by postselection of the APD signal affects predominately the first mode. The effect of photon subtraction on other modes can be disregarded since the subtraction probability
is proportional to $\sinh^2r_j$. Therefore, approximating a photon-subtracted single-mode squeezed state as a single-photon state, we can evaluate the TRWF value at the origin as $\frac{1}{\pi}\big[1-2T_1(t_d)\big]$,
which up to the prefactor $1/\pi$ is the mean value of the parity operator for $\hat{A}^{(d)}(t_d)$ (cf. ref.~\cite{Vogel2006}, p.~123). The quantum nature becomes pronounced in both the metrological power and the negativity unless the probe pulse duration is too short so that the results are strongly contaminated by the vacuum contributions arising effectively from high-order principal modes. Expanding the probe pulse duration to $t_p\approx24.5\,$fs, we get the highest values of the metrological power 
and negativity 
reflected in the TRWF (Figs.~\ref{Fig:MPandNegativity}c and 4d). A larger duration leads to decreases in these quantities.

In summary, we have developed a theory of the time-domain subcycle tomography of ultrabroadband quantum light
based on a variant of the BHD harnessing
ultrashort phase-controlled LOs or alternatively on EOS.
In particular, we demonstrated how to reconstruct phase-space distributions of ultrashort Gaussian and non-Gaussian states of pulsed quantum light with a resolution corresponding to the duration of the probe pulse, which is selected sufficiently below the characteristic timescale of the studied signal.
For that, we generalized the definition of the usual mode-fixed Wigner function
by introducing the TRWF capturing the subcycle evolution of the states.
Analyzing this evolution, we can confirm
the description of the squeezing process in the time domain as an effective modulation of the time flow experienced by the quantum field in the generation process.
In order to get a deeper insight into the dynamics of the TRWF and to connect it to the mode analysis commonly exploited
in quantum optics, we
decomposed the sampled pulsed squeezed state into principal modes through the Bloch-Messiah reduction. We found that the TRWF is mainly determined by a superposition of just few of these \textit{a priori} unknown modes whereas the rest of the modes can be seen as an effective vacuum. The dynamics of the TRWF, which shows the pulsed squeezing, follows from the squeezing parameters of the dominant principal modes and their time-dependent overlap with the effective detection mode.
%
Whereas a higher temporal resolution is achieved by reducing the
probe pulse duration, the phase-space resolution required to capture the quantum nature of the states  becomes compromised
by contributions from higher-order modes stemming from the background vacuum. Nevertheless, the quantum features are pronounced even if the probe pulse duration
is still considerably smaller than the characteristic period of
the field driving the quantum light.
Time-dependent non-Gaussianity generated by the single-photon subtraction, realized via the beam splitter with the APD, can also be observed in the TRWF. Our subcycle tomography shows  negativity in the TRWF of the photon-subtracted state, even without an exact mode matching. This fact renders our
protocol attractive for applications in quantum metrology and quantum computing.
Such remarkable capabilities to trace quantum states of light with an unprecedented temporal resolution open new horizons in ultrafast quantum photonics.
In particular, \textit{a priori} unknown temporal modes of ultrabroadband quantum light
can be reconstructed from the observed TRWF (see Supplementary Information). On the technical side, we like to highlight that the reconstruction based on
the Gram-Charlier expansion shall significantly reduce the number of the measurements required for a desired precision. This might be a vital feature, since millions of repetitions are required in the relevant state-of-the-art experiments to appropriately capture the full quantum statistics of just one field quadrature at each time moment \cite{THz_Roadmap2023}.
%
%
Finally, the presented subcycle tomography can provide time-domain access to  ultrabroadband quantum light bearing a potential to become a new-generation tool to probe ultrafast quantum phenomena, such as a rapid build-up of entanglement which is inaccessible by classical light.

%


\setcounter{section}{0}

\section*{Methods}

\subsection{Generation of the pulsed squeezed vacuum}\label{Methods:Generation}
We assume that the nonlinear interaction in the GX is driven by a strong coherent field, with a particular temporal profile given by
$\mathcal{E}(t)=\mathcal{E}_0 \, \sech(\Gamma t)$
with $\Gamma=2 \mathrm{arcsech}(\frac{1}{2}) / \delta_d$, where the full width at half maximum (FWHM) is $\delta_d=16$~fs. As shown in ref.~\cite{KizmannSubcycle}, the quantum field generated by the interaction with the co-propagating vacuum field $\hat{E}_\mathrm{vac}$ can be characterized by an effective squeezing strength $r_\mathrm{eff}$, which is proportional to the nonlinear coefficient and the length of the GX, $\mathcal{E}_0$, and $1/\delta_d$.  The temporal evolution of this quantum field can be directly captured by the so-called conformal time $\tau(t)$,  $\hat{E}(t)=\tau'(t)\hat{E}_\mathrm{vac}\big(\tau(t)\big)$ with $\tau'(t)$ denoting here the temporal derivative. The input and output electric fields at the GX are connected then in the frequency domain by the corresponding Bogoliubov transformation \cite{KizmannSubcycle}: $\hat{b}(\omega)=\int_0^\infty\mathrm{d}\omega'\, p(\omega,\omega')\hat{a}(\omega')+\int_0^\infty\mathrm{d}\omega'\, q(\omega,\omega')\hat{a}^\dagger(\omega')$ with $p(\omega,\omega')=-q(\omega,-\omega')=\frac{1}{2\pi}\sqrt{\big\lvert \frac{\omega'}{\omega}\big\rvert}\int_{-\infty}^\infty \mathrm{d}t\allowbreak\, e^{i\omega \tau^{-1}(t)-i\omega' t}$. Here, $\hat{a}$ ($\hat{b}$) is the annihilation operator of the input (output) field and $\tau^{-1}(t)$ denotes the inverse function of $\tau(t)$. The indicated shape of the driving field leads to
$\tau(t)=\frac{1}{\Gamma} \mathrm{arcsinh}\left(\sinh(\Gamma t)+r_\mathrm{eff}\right)$. We select $r_\mathrm{eff}=5$ for the majority of our calculations.

\subsection{Sampling of the electric-field quadratures}\label{Methods:Sampling}
The subcycle tomography measurements aim to construct a quasiprobability distribution of the positive-frequency part of the electric-field operator $\hat{E}^{(+)}(t)=\frac{1}{2} \big[\hat{E}(t)-i\hat{E}_{\frac{\pi}{2}}(t)\big]$. Since the diverging commutation relation $\big[\hat{E}(t),\hat{E}_{\frac{\pi}{2}}(t)\big]=-2iC\int_{0}^{\infty}\mathrm{d}\omega\, \omega$, where $C$ is a constant, prevents a definition of generalized field quadratures, which would be appropriately normalized, we resort
%
to a scheme shown in Fig.~\ref{Fig:Subcycle Tomography by BHD}b, where the field $\hat{E}^{(+)}(t)$ is sampled with a given temporal resolution $\delta_p$ leading to $\hat{E}^{(d)}_{\phi}(t_d)=\int_{-\infty}^{\infty} \mathrm{d}t \, R(t-t_d)\hat{E}_{\phi}(t)\equiv(R\star \hat{E}_{\phi})(t_d)$ for all values of $\phi$. Here $R(t)$ is an appropriate gating function determined by the temporal profile of the LO. In our calculations we assume that it is given by a Gaussian $R(t)=\frac{2 \sqrt{\ln 2}}{\sqrt{\pi}\delta_p} e^{-\frac{4\ln 2 \,t^2}{\delta_p^2}}$, where $\delta_p$ represents the FWHM. Then, the commutation relation for the \textit{detected} field leads to a finite value $\big[\hat{E}^{(d)}(t_d),\hat{E}^{(d)}_{\frac{\pi}{2}}(t_d)\big]=-2iC\frac{\text{ln}16}{\delta_p^2}$.
Since a CEP shift applied on the field gives $\hat{E}^{(d)}_{\phi}(t)=\hat{E}^{(d)}(t) \cos \phi+\hat{E}^{(d)}_{\frac{\pi}{2}}(t) \sin \phi$, we can select $\hat{E}^{(d)}(t)$ and $\hat{E}^{(d)}_{\frac{\pi}{2}}(t)$ as (not yet normalized) quadratures. The normalization is assured by dividing $\hat{E}^{(d)}(t)$ and $\hat{E}^{(d)}_{\frac{\pi}{2}}(t)$ by $\sqrt{\big\lvert \big[\hat{E}^{(d)}(t_d),\hat{E}^{(d)}_{\frac{\pi}{2}}(t_d)\big]\big\rvert}$, as  in equation \eqref{eq:quadratures}. Hence, the described subcycle tomography protocol corresponds to a reconstruction of the Wigner function of the detection mode $\hat{A}^{(d)}(t_d)=-i\sqrt{2}\mathcal{N}(R\star\hat{E}^{(+)})(t_d)$, so that $W(x,p;t_d)$ captures $\hat{E}^{(+)}(t)$ with resolution $\delta_p$ at the time moment $t_d$ .

In the case of a separable state such as the pulsed squeezed state, modal decomposition of the positive-frequency part of the electric-field operator  $\hat{E}^{(+)}(t)$ with respect to the principal modes $\{\hat{b}_j\}$, i.e. $\hat{E}^{(+)}(t)=\sum_j \alpha_j(t)\,\hat{b}_j
$\cite{Raymer2020}, leads to
\begin{eqnarray}\label{eq:modaldecompositionofA}
\hat{A}^{(d)}(t_d)=\sum_j \theta_j(t_d)\hat{b}_j
\end{eqnarray}
with $\theta_j(t_d)$ given by equation \eqref{eq:transmittance}.
According to equation \eqref{eq:modaldecompositionofA}, the observed quantum nature of the sampled light originates from the states of $\{\hat{b}_j\}$ whereas the temporal dynamics can be attributed to the field modes $\{\alpha_j(t)\}$ resolved by the probe pulse $R(t)$.

\subsection{The Gram-Charlier expansion}
Knowing the lowest statistical moments to a certain order, the probability distribution can be inferred with a corresponding precision through the Gram-Charlier expansion. Within this method, an unknown probability distribution $f(x_1,x_2)$ of two variables, as we need in our case, is expressed via a given probability distribution $g(x_1,x_2)$ as\cite{Sauer1979}
\begin{eqnarray}\label{eq:GCexpansion}
f(x_1,x_2)=\text{exp}\big[\sum_{j,k}(K_{j,k}-\gamma_{j,k})\frac{(-D_{x_1})^{j}(-D_{x_2})^{k}}{j!k!}\big]g(x_1,x_2)\;.
\end{eqnarray}
Here $D_{x_i}$ represents the partial derivative with respect to $x_i$ ($i=1,2$), whereas $K_{j,k}$ and $\gamma_{j,k}$ are the cumulants of $f(x_1,x_2)$ and $g(x_1,x_2)$, respectively. Such expansion
is also valid for the quasiprobability distributions with the cumulants recovered from the corresponding moments\cite{Cohen1998}. In the case of the TRWF, the moments stand for the mean values of the symmetrically ordered moments of $\hat{X}(t_d)$ and $\hat{P}(t_d)$ (for instance, the 2nd-order moments comprise $\langle\hat{X}^2(t_d)\rangle$, $\langle\hat{P}^2(t_d)\rangle$, and $\big\langle\frac{\hat{X}(t_d)\hat{P}(t_d)+\hat{P}(t_d)\hat{X}(t_d)}{2}\big\rangle$).
The reference $g(x_1,x_2)$ is typically chosen as a normal distribution.
Here we choose the reference to represent the vacuum, $W_\mathrm{vac}(x,p)=\frac{1}{\pi}e^{-x^2-p^2}$, which we have initially for the analyzed light state. Then equation \eqref{eq:GCexpansion} reads
\begin{eqnarray}\label{eq:GCexpansionofW}
W(x,p;t_d)=\sum_{n,m}C_{nm}H_n(x)H_m(p)W_\mathrm{vac}(x,p),
\end{eqnarray}
where $H_n(x)$ denote the $n$th-order Hermite polynomials and the coefficients $C_{nm}$ are given by $C_{nm}=\frac{1}{2^{n+m}n!m!}\iint \mathrm{d}x \mathrm{d}p H_n(x) H_m(p) \allowbreak W(x,p;t_d)$. Unlike the case of probability distributions, we do not need to concern about negative values in particular contributions to $W(x,p,t_d)$ coming out from the Hermite polynomials. Note that each of the coefficients $C_{nm}$ can be expressed via a linear combination of the mean values of the symmetrically ordered moments of $\hat{X}(t_d)$ and $\hat{P}(t_d)$ up to the (\textit{n}+\textit{m})th order (see Supplementary Information).

The efficiency of equation \eqref{eq:GCexpansionofW} for approximating the TRWF can be evaluated from a number of the required measurements to implement the expansion to a certain order.
To reconstruct $W(x,p;t_d)$ from the inverse Radon transform, one has to acquire statistics of the phase-rotated quadratures for every value of the CEP shift, or in practice for a sufficiently large number of values to perform numerical integration,  whereas the Gram-Charlier expansion relies on a comparatively low number of symmetrized moments of $\hat{X}(t_d)$ and $\hat{P}(t_d)$. It also does not need to care about the kernel divergence problems inherent to the Radon transform.  Approximating the TRWF up to the \textit{N}th order of equation \eqref{eq:GCexpansionofW},  $m+n\leq N$, requires the moments till the $N$th order. As the number of these moments is $N+1$ (number of the elements in the corresponding row of Pascal's triangle), to obtain their values, measurements for at least $N+1$ different CEP shifts, defining the corresponding generalized quadratures, have to be performed. For each of the shifts, only the lowest $N$ moments of these quadratures must be acquired, not necessarily the whole statistical distribution.  In practice this means a lower number of  repetitions should be sufficient to get a specific signal-to-noise ratio.  The resulting linear system of $N+1$ equations for $N+1$ variables is solvable (see Supplementary Information). The solution is used in equation \eqref{eq:GCexpansionofW}, resulting in the reconstructed TRWF for the given $t_d$.

\section*{Acknowledgements}
This research was supported by the National Research Foundation of Korea (NRF) grant funded by the Korea government (MSIT) (2020R1A2C1008500). M.K., A.L. and A.S.M. acknowledge funding by the Deutsche Forschungsgemeinschaft (DFG)--Project No. 425217212--SFB 1432. M.K. gratefully acknowledges the support from the Alexander von Humboldt Foundation
through the Feodor Lynen program.



\renewcommand{\theequation}{S\arabic{equation}}
\renewcommand{\thefigure}{S\arabic{figure}}
\setcounter{equation}{0}
\setcounter{figure}{0}
\setcounter{section}{0}

\begin{flushleft}
\textbf{\large Supplementary Information}
\end{flushleft}


\section{Scheme based on electro-optic sampling}

\begin{figure}[h]
    \includegraphics[width=16cm]{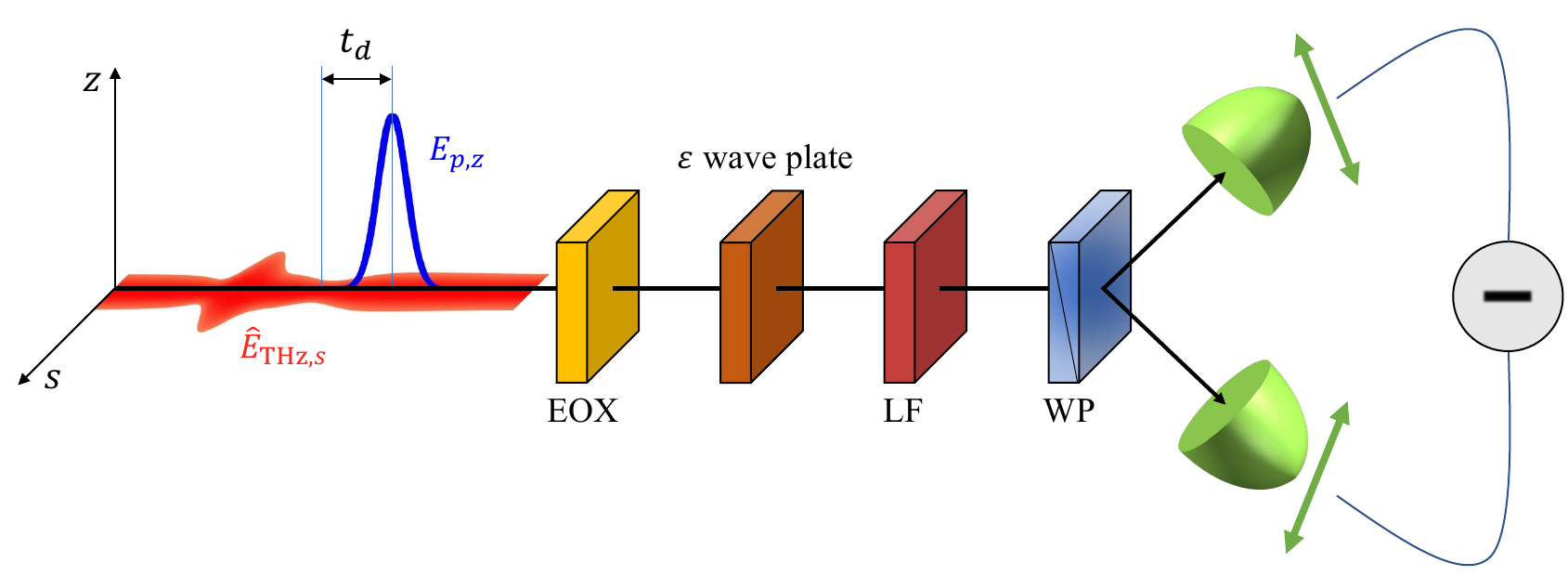}
    \centering
    \caption{
    \textbf{Schematic of subcycle tomography setup based on EOS.} The incoming quantum light $\hat{E}_{\mathrm{THz},s}$ (red band) interacts with a co-propagating probe pulse $E_{p,z}$ (blue) centered at a variable time delay $t_d$ in an electro-optic crystal (EOX). The interaction induces a change in the polarization state of the probe, which is analyzed by the following adjustable ellipsometer, consisting of a wave plate, a Wollaston prism (WP) and two balanced photodetectors receiving probe photons with mutually perpendicular polarization. Longpass filter (LF) with frequency response $\mathcal{F}(\omega)$ is used to attenuate high frequency of the probe pulse. Tuning the phase shift $\varepsilon$
    induced by the wave plate, supplemented by its appropriate rotation, allows to get access to quantum statistics of all detected phase-shifted field operators $\hat{E}^{(d)}_\phi(t_d)$ \cite{Sulzer2020,Kizmann2022,Unruh-DeWitt}, where $\phi=\phi(\varepsilon)$ as described in the text.
    \label{EOS}}
    %
\end{figure}

In this section, we show that electro-optic sampling (EOS)\cite{Riek2015,Benea2019,Sulzer2020,Kizmann2022} is analogous to the pulsed BHD described in the main text. EOS is realized by the setup shown in Fig.~\ref{EOS}. A THz quantum field $\hat{E}_{\mathrm{THz},s}$ is sent into a second-order nonlinear crystal together with an ultrashort NIR probe pulse $E_{\mathrm{p},z}$. Inside the crystal, the quantum statistics of the THz field are imprinted on the quantum statistics of the polarization state of the probe by generating a weak NIR field along the polarization perpendicular to the probe $\hat{E}_{\mathrm{p},s}$. The ellipsometry, consisting of a wave plate and a polarization dependent beam splitter, is then used to measure the statistics of the polarization state of the probe. Measuring the statistics of the polarization state of the probe for different time delays reveals then the time-dependent statistics of the incoming THz quantum field.

The electro-optic signal in Fig.~\ref{EOS} is given by\cite{Sulzer2020,Kizmann2022}
\begin{align}
\hat{\mathcal{S}}(\phi)=C'\int_0^\infty\!\!\! \mathrm{d}\omega\; \frac{\mathcal{F}(\omega)}{\hbar\omega}\left[e^{i\phi}E^*_{\mathrm{p},z}(\omega)\hat{E}_{\mathrm{p},s}(\omega)+H.c.\right].\label{signal2}
\end{align}
Here, $C'=4\pi\varepsilon_0 n A c_0$
with $A$ as the effective transverse area determined by the beam waist of the probe field, $n$ refractive index of the EOM for the probe pulse, $c_0$ the speed of light in vacuum, and $\varepsilon_0$ the vacuum permittivity. The phase shift $\phi=\arccos(\sqrt{-\cos(\varepsilon)})$ is obtained by changing the phase shift $\varepsilon$ of the wave plate with $\varepsilon\in\{\pi/2,3\pi/2\}$. $\mathcal{F}(\omega)$ is the spectral filtering function, which describes the case when only a part of the spectrum of the probe is collected.

This signal can also be rewritten in terms of a (not normalized) multimode annihilation operator $\hat{\mathfrak{a}}(\phi)$,
\begin{align}
\hat{\mathcal{S}}(\phi)=\hat{\mathfrak{a}}(\phi)+\hat{\mathfrak{a}}^\dagger(\phi),\label{signal3}
\end{align}
where
\begin{align}
\hat{\mathfrak{a}}(\phi)=\sqrt{N}e^{i\phi}\int_0^\infty\!\!\mathrm{d}\omega~ h(\omega)\hat{a}_s(\omega).\label{b}
\end{align}
Here, $N=C'\int_0^\infty\!\!\mathrm{d}\omega |\mathcal{F}(\omega)E_{\mathrm{p},z}(\omega)|^2/(\hbar\omega)$ is the mean number of photons in the spectrally filtered probe, corresponding also to the shot-noise level, and $\hat{a}_s(\omega)$ annihilates an $s$-polarized photon of frequency $\omega$. In order to match equation \eqref{signal3} to equation \eqref{signal2}, we define the ultrabroadband frequency mode
\begin{align}
h(\omega)=i\frac{\mathcal{F}(\omega)E^*_{\mathrm{p},z}(\omega)/\sqrt{\omega}}{\sqrt{\int_0^\infty\!\!\mathrm{d}\omega |\mathcal{F}(\omega)E_{\mathrm{p},z}(\omega)|^2/\omega}},\label{h}
\end{align}
such that $\int_0^\infty \!\!\mathrm{d}\omega|h(\omega)|^2=1$ and therefore $\left[\hat{\mathfrak{a}}(\phi),\hat{\mathfrak{a}}^\dagger(\phi)\right]=N$.
The signal can thus be alternatively understood as a realization of a homodyne detection of the $s$-polarized state according to a local oscillator mode $h(\omega)$ essentially given by the probe $E_{\mathrm{p},z}$. This means that the ellipsometry scheme in Fig.~\ref{EOS}, consisting of a waveplate, a Wollaston prism and two detectors, can be regarded as equivalent to a homodyne detection setup, where the waveplate is used to control the phase of the local oscillator, the Wollaston prism replaces the beam splitter, and the two directions of polarization 
act as the two input ports of the beam splitter. Note that due to the fact that in EOS, the carrier frequency of the probe can be much higher as compared to the signal frequency, it is practically easier in this implementation to reach subcycle temporal resolution.

In the case of EOS, we are specifically interested in the NIR field generated through the interaction of the probe with a given THz quantum field $\hat{E}_{\mathrm{THz},s}$  inside the electro-optic crystal. The goal is then to reconstruct the statistics of $\hat{E}_{\mathrm{THz},s}$ from the statistics of the generated NIR field $
\hat{E}_{p,s}$. The probability distribution of the electro-optic signal is given by\cite{Kizmann2022}
\begin{align}
\begin{split}
P(\mathcal{S},\phi)&=\frac{1}{\sqrt{2\pi N}}\langle 0_\mathrm{NIR} ,\Psi_\mathrm{THz}|\hat{U}^\dagger :\! e^{-\frac{\left(\mathcal{S}-\hat{\mathcal{S}}(\phi)\right)^2}{2N}}\!:\hat{U}|\Psi_\mathrm{THz},0_\mathrm{NIR}\rangle\\
&=\frac{1}{\sqrt{2\pi N}}\sum_{k=0}^\infty \frac{1}{(2N)^{k/2} k!}H_k(\mathcal{S}/\sqrt{2N})\exp\left(-\frac{\mathcal{S}^2}{2N}\right)
\langle 0_\mathrm{NIR} ,\Psi_\mathrm{THz}|\hat{U}^\dagger:\! \hat{\mathcal{S}}^k(\phi)\! :\hat{U}|\Psi_\mathrm{THz},0_\mathrm{NIR}\rangle,
\end{split}\label{prob2}
\end{align}
with $H_k(x)$ being the Hermite polynomials of order $k$ and $:\boldsymbol{\cdot}:$ indicating normal ordering. Here, the evolution operator $\hat{U}$ is used to describe the electro-optic process inside the nonlinear crystal and $|\Psi_\mathrm{THz},0_\mathrm{NIR}\rangle=|\Psi_\mathrm{THz}\rangle\otimes |0_\mathrm{NIR}\rangle$ denotes the initial (separable) state at the entrance of the crystal, consisting of the state of the sampled $s$-polarized THz quantum field and the $s$-polarized NIR vacuum state, respectively.

The electro-optic process results from both difference-frequency generation (DFG), which couples annihilation and creation operators, and sum-frequency generation (SFG), coupling annihilation operators of different frequencies. Therefore, the evolution of $\hat{\mathfrak{a}}$ under $\hat{U}$ can be described in general by a Bogoliubov transformation of the form
\begin{align}
\hat{U}^\dagger\hat{\mathfrak{a}}(\phi)\hat{U}=\hat{\alpha}(\phi)+\hat{\beta}^\dagger(\phi),\label{btrans}
\end{align}
where we have separated the operator $\hat{U}^\dagger\hat{\mathfrak{a}}(\phi)\hat{U}$ into its annihilation
\begin{align}
\hat{\alpha}(\phi)=\sqrt{N}e^{i\phi}\iint_0^\infty\!\!\mathrm{d}\omega \mathrm{d}\overline{\omega} ~h(\omega)p(\omega,\overline{\omega})\hat{a}_s(\overline{\omega})\label{alpha}
\end{align}
and creation
\begin{align}
\hat{\beta}^\dagger(\phi)=\sqrt{N}e^{i\phi}\iint_0^\infty\!\!\mathrm{d}\omega \mathrm{d}\overline{\omega} ~h(\omega)q(\omega,\overline{\omega})\hat{a}_s^\dagger(\overline{\omega})\label{beta}
\end{align}
parts, such that $\hat{\alpha}(\phi)|0\rangle=0$ and $\hat{\beta}(\phi)|0\rangle=0$. Here, we use the most general form of the Bogoliubov transformation, which means that the only restriction to the operators $\hat{\alpha}(\phi)$ and $\hat{\beta}(\phi)$ is
\begin{align}
\left[\hat{\mathfrak{a}}(\phi),\hat{\mathfrak{a}}^\dagger(\phi)\right]=\left[\hat{U}^\dagger \hat{\mathfrak{a}}(\phi)\hat{U},\hat{U}^\dagger\hat{\mathfrak{a}}^\dagger(\phi)\hat{U}\right]=\left[\hat{\alpha}(\phi),\hat{\alpha}^\dagger(\phi)\right]-\left[\hat{\beta}(\phi),\hat{\beta}^\dagger(\phi)\right]=N,
\end{align}
in accordance to $\int_0^\infty\mathrm{d}\omega~\left[p(\omega_1,\omega)p^*(\omega_2,\omega)-q(\omega_1,\omega)q^*(\omega_2,\omega)\right]=\delta(\omega_1-\omega_2)$. Note that the operator $\hat{\mathfrak{a}}(\phi)$ in equation \eqref{b} acts only on the NIR frequencies due to the selected broadband mode $h(\omega)$ of the probe, given in equation \eqref{h}. In contrast, the transformed operator $\hat{U}^\dagger\hat{\mathfrak{a}}(\phi)\hat{U}$ in equation \eqref{btrans} acts on both the NIR and THz frequencies, since the Bogoliubov transformation couples these two frequency ranges through the functions $p(\omega,\overline{\omega})$ and $q(\omega,\overline{\omega})$ in equations \eqref{alpha} and \eqref{beta}.

We can now use this Bogoliubov transformation to bring the operator $\hat{U}^\dagger:\! \hat{\mathcal{S}}^k(\phi)\! :\hat{U}$ in the second line of equation \eqref{prob2} into its normally ordered form
\begin{align}
\begin{split}
\hat{U}^\dagger:\! \hat{\mathcal{S}}^k\! :\hat{U}&=\sum_{l=0}^k \binom{k}{l}(\hat{\alpha}^\dagger+\hat{\beta})^{k-l}(\hat{\alpha}+\hat{\beta}^\dagger)^l\\
&=\sum_{l=0}^k \binom{k}{l}\sum_{i,j=0}^{k-l}\sum_{s,t=0}^l I^{(k-l)}_{i,j}\left([\hat{\beta},\hat{\alpha}^\dagger]\right)I^{(l)}_{s,t}\left([\hat{\alpha},\hat{\beta}^\dagger]\right)\hat{\alpha}^{\dagger i}\hat{\beta}^j\hat{\beta}^{\dagger s}\hat{\alpha}^t\\
&=\sum_{l=0}^k \binom{k}{l}\sum_{i,j=0}^{k-l}\sum_{s,t=0}^l \sum_{r=0}^{\mathrm{min}(j,s)} I^{(k-l)}_{i,j}\left([\hat{\beta},\hat{\alpha}^\dagger]\right)I^{(l)}_{s,t}\left([\hat{\alpha},\hat{\beta}^\dagger]\right)\binom{j}{r}\binom{s}{r}
r![\hat{\beta},\hat{\beta}^\dagger]^r\hat{\alpha}^{\dagger i}\hat{\beta}^{\dagger s-r}\hat{\beta}^{j-r}\hat{\alpha}^t,
\end{split}\label{normordS}
\end{align}
where we have omitted the dependencies on $\phi$. In the first line, we have used the unitarity $\hat{U}\hat{U}^\dagger=1$ of the operator $\hat{U}$ to transform both $\hat{\mathfrak{a}}$ and $\hat{\mathfrak{a}}^\dagger$. In the second line, we used the formula $(\hat{\alpha}+\hat{\beta}^\dagger)^l=\sum_{s,t=0}^l I^{(l)}_{s,t}\left([\hat{\alpha},\hat{\beta}^\dagger]\right) \hat{\beta}^{\dagger s}\hat{\alpha}^t$, with
\begin{align}
I^{(l)}_{s,t}\left(x\right)=\begin{cases*}
      \frac{l!}{2^{\frac{l-s-t}{2}}\left(\frac{l-s-t}{2}\right)! s! t!} x^\frac{l-s-t}{2} & if $l-s-t$ even; \\
      0       & if $l-s-t$ odd,
    \end{cases*}
\end{align}
 which can be derived by equating coefficients in the Baker-Campbell-Hausdorff formula in the form $e^{z(\hat{\alpha}+\hat{\beta}^\dagger)}=e^{z\hat{\beta}^\dagger}e^{z\hat{\alpha}}e^{z^2[\hat{\alpha},\hat{\beta}^\dagger]/2}$. In the last line of equation \eqref{normordS}, we have transformed the operator $\hat{\beta}^j\hat{\beta}^{\dagger s}$ into its normally ordered form, which can also be done by equating coefficients in the Baker-Campbell-Hausdorff formula in the form $e^{z\hat{\beta}}e^{z\hat{\beta}^\dagger}=e^{z\hat{\beta}^\dagger}e^{z\hat{\beta}}e^{z^2[\hat{\beta},\hat{\beta}^\dagger]}$.

 Equation~\eqref{normordS} can now be simplified further by first combining the annihilation operators and then the creation operators. The creation and annihilation operators can then also be combined using normal ordering
 as
 \begin{align}
 \hat{U}^\dagger:\! \hat{\mathcal{S}}^k\! :\hat{U}&=\sum_{m=0}^k \binom{k}{m} (k-m-1)\tilde{!!} \kappa^\frac{k-m}{2} :\left(\hat{\alpha}+\hat{\beta}+\hat{\alpha}^\dagger+\hat{\beta}^\dagger\right)^m:,\label{normordS2}
 \end{align}
 where $(k-m-1)\tilde{!!}=\frac{(k-m)!}{2^\frac{k-m}{2} \frac{k-m}{2}!}$ if $k-m$ is even and $(k-m-1)\tilde{!!}=0$ if $k-m$ is odd (notice here the difference to the usual double factorial $!!$ notation for the odd case). Here,
 \begin{align}
 \kappa\equiv\kappa(\phi)=2\left[\hat{\beta}(\phi),\hat{\beta}^\dagger(\phi)\right]+\left(\left[\hat{\alpha}(\phi),\hat{\beta}^\dagger(\phi)\right]+H.c.\right)=\langle 0_{\mathrm{NIR}},0_{\mathrm{THz}}|\hat{U}^\dagger \!:\!\hat{\mathcal{S}}^2(\phi)\!:\! \hat{U}|0_{\mathrm{THz}},0_{\mathrm{NIR}}\rangle\label{kappa}
 \end{align}
is given by the normally ordered second moment of the electro-optic signal for $|\Psi_{\mathrm{THz}}\rangle=|0_{\mathrm{THz}}\rangle$ and the corresponding normally ordered $k$th-moment is given by $(k-1)\tilde{!!}\kappa^\frac{k}{2}(\phi)$. Note that the commutator $\left[\hat{\alpha}(\phi),\hat{\alpha}^\dagger(\phi)\right]$ does not contribute to $\kappa(\phi)$, while the commutator $\left[\hat{\beta}(\phi),\hat{\beta}^\dagger(\phi)\right]$ contribute twice. This is a result of the normal ordering in equation \eqref{prob2}. The operator $\hat{\alpha}(\phi)$ can be essentially identified as the SFG contribution, which does not generate any photons from the vacuum. Correspondingly, $\hat{\beta}(\phi)$ is identified with the DFG contribution, which generates both NIR and THz photons.

 Inserting equation \eqref{normordS2}
 into equation \eqref{prob2}, we finally obtain
\begin{align}
\begin{split}
P(\mathcal{S},\phi)=\frac{1}{\sqrt{2\pi [N+\kappa(\phi)]}}\langle 0_\mathrm{NIR} ,\Psi_\mathrm{THz}|\! :\!\exp\left(-\frac{\left(\mathcal{S}-\hat{U}^\dagger\hat{\mathcal{S}}(\phi)\hat{U}\right)^2}{2(N+\kappa(\phi))}\right)\! :\!|\Psi_\mathrm{THz},0_\mathrm{NIR}\rangle.
\end{split}\label{prob3a}
\end{align}
Equation~\eqref{prob3a} looks very similar to the first line of equation \eqref{prob2}, however, we have replaced the shot noise contribution $N=\langle 0_\mathrm{NIR}|\hat{\mathcal{S}}^2(\phi)|0_{\mathrm{NIR}}\rangle$ by $N+\kappa(\phi)=\langle 0_{\mathrm{NIR}},0_{\mathrm{THz}}|\hat{U}^\dagger \hat{\mathcal{S}}^2(\phi)\hat{U}|0_{\mathrm{THz}},0_{\mathrm{NIR}}\rangle$ and have shifted the evolution operators inside
the normal ordering operation $:\boldsymbol{\cdot}:$. As a result, the transformed electro-optic signal $\hat{U}^\dagger\hat{\mathcal{S}}(\phi)\hat{U}=\hat{\alpha}(\phi)+\hat{\beta}(\phi)+H.c.$ now not only acts on the $s$-polarized NIR state $|0_{\mathrm{NIR}}\rangle$, but also on the THz state $|\Psi_\mathrm{THz}\rangle$. Therefore, the electro-optic process preserves the Gaussian properties of incoming states, so that, e.g., for $|\Psi_\mathrm{THz}\rangle=|0\rangle$ the initial variance $N$ in equation \eqref{prob2} is simply changed to the variance $N+\kappa(\phi)$ in equation \eqref{prob3a}. This is not surprising since it results from a mixture of SFG and DFG processes, and squeezing processes are known to have this property \cite{Schumaker1986}.

The probability distribution in equation \eqref{prob3a} is given by a normally ordered operator and the $s$-polarized NIR field component is in its ground state $|0_{\mathrm{NIR}}\rangle$. Thus, it is easy to see that we only need to take into account the part of the transformed electro-optic signal $\hat{U}^\dagger\hat{\mathcal{S}}(\phi)\hat{U}$ that acts on the THz state. In order to do so, we split the transformed electro-optic signal into the NIR $\hat{\tilde{\mathcal{S}}}_{\mathrm{NIR}}(\phi)$ and THz $\hat{\tilde{\mathcal{S}}}_{\mathrm{THz}}(\phi)$ parts:
\begin{align}
\hat{U}^\dagger\hat{\mathcal{S}}\hat{U}=\hat{\tilde{\mathcal{S}}}_{\mathrm{NIR}}+\hat{\tilde{\mathcal{S}}}_{\mathrm{THz}}=\left(\hat{\alpha}_{\mathrm{NIR}}+\hat{\beta}_{\mathrm{NIR}}+H.c.\right)+\left(\hat{\alpha}_{\mathrm{THz}}+\hat{\beta}_{\mathrm{THz}}+H.c.\right),\label{transfS}
\end{align}
where we have again omitted the $\phi$ dependence for brevity. Here, the operators $\hat{\alpha}_{\mathrm{NIR}}(\phi)$ and $\hat{\alpha}_{\mathrm{THz}}(\phi)$ are defined as
\begin{subequations}
\begin{align}
\hat{\alpha}_{\mathrm{NIR}}(\phi)&=\sqrt{N}e^{i\phi}\int_0^\infty\!\!\mathrm{d}\omega \int_{\Omega_{\mathrm{max}}}^\infty\mathrm{d}\overline{\omega} ~h(\omega)p(\omega,\overline{\omega})\hat{a}_s(\overline{\omega})\label{alphaNIR}\\
\hat{\alpha}_{\mathrm{THz}}(\phi)&=\sqrt{N}e^{i\phi}\int_0^\infty\!\!\mathrm{d}\omega \int_0^{\Omega_{\mathrm{max}}}\mathrm{d}\Omega ~h(\omega)p(\omega,\Omega)\hat{a}_s(\Omega)\label{alphaTHz},
\end{align}
\end{subequations}
where we have used the integration variables $\overline{\omega}$ and $\Omega$ to portray the fact that $\hat{\alpha}_{\mathrm{NIR}}(\phi)$ acts on the NIR frequencies $\overline{\omega}$ and $\hat{\alpha}_{\mathrm{THz}}(\phi)$ acts on the THz frequencies $\Omega$. The operators $\hat{\beta}_{\mathrm{NIR}}(\phi)$ and $\hat{\beta}_{\mathrm{THz}}(\phi)$ can be obtained in an equivalent way to equations \eqref{alphaNIR} and \eqref{alphaTHz}, respectively. The frequency $\Omega_{\mathrm{max}}$ denotes the boundary between the NIR and THz frequencies and should be chosen with respect to the spectral width of $h(\omega)$ in equation \eqref{h}, such that $h(\Omega_{\mathrm{max}})\ll 1$. We can now calculate the expectation value with respect to the NIR state $|0_{\mathrm{NIR}}\rangle$, which is given by
\begin{align}
\begin{split}
P(\mathcal{S},\phi)=\frac{1}{\sqrt{2\pi [N+\kappa(\phi)]}}\langle \Psi_\mathrm{THz}|\! :\!\exp\left(-\frac{\left(\mathcal{S}-\hat{\tilde{\mathcal{S}}}_\mathrm{THz}(\phi)\right)^2}{2[N+\kappa(\phi)]}\right)\! :\!|\Psi_\mathrm{THz}\rangle.
\end{split}\label{prob3}
\end{align}
We see that the statistics of the electro-optic signal are indeed determined by the statistics of the sampled THz field.

The operator inside the expectation value of equation~\eqref{prob3} is effectively given by a Gaussian function. The fact that this Gaussian function is normally ordered in fact means that we can treat the operator $\hat{\tilde{\mathcal{S}}}_{\mathrm{THz}}(\theta)$ in its exponent as basically a normal $c$-number instead of an operator, since annihilation and creation operators commute within a normally ordered product $:\hat{a}\hat{a}^\dagger:=:\hat{a}^\dagger\hat{a}:=\hat{a}^\dagger \hat{a}$ \cite{Fan2003}. This allows us to write equation \eqref{prob3} as a convolution of two probability distributions $P'(\mathcal{S},\phi)$ and $P_{\mathrm{THz}}(\mathcal{S},\phi)$:
\begin{align}
P(\mathcal{S},\phi)=\int_{-\infty}^\infty\!\!\mathrm{d}\mathcal{S}' P'(\mathcal{S}-\mathcal{S}',\phi)P_{\mathrm{THz}}(\mathcal{S}',\phi).\label{conv}
\end{align}
The probability distribution $P'(\mathcal{S},\phi)$ is given by
\begin{align}
P'(\mathcal{S},\phi)=\frac{1}{\sqrt{2\pi \tilde{\kappa}(\phi)}}\exp\left(-\frac{\mathcal{S}^2}{2\tilde{\kappa}(\phi)}\right),\label{prob'}
\end{align}
where $\tilde{\kappa}=N+2\left[\hat{\beta}_{\mathrm{NIR}},\hat{\beta}_{\mathrm{NIR}}^\dagger\right]+\left(\left[\hat{\alpha}_{\mathrm{NIR}},\hat{\beta}_{\mathrm{NIR}}^\dagger\right]+H.c.\right)-\left[\hat{\alpha}_{\mathrm{THz}},\hat{\alpha}_{\mathrm{THz}}^\dagger\right]+\left[\hat{\beta}_{\mathrm{THz}},\hat{\beta}_{\mathrm{THz}}^\dagger\right]$. This means that the probability distribution $P_{\mathrm{THz}}(\mathcal{S},\phi)$ is given by
\begin{align}
\begin{split}
P_{\mathrm{THz}}(\mathcal{S},\phi)=\frac{1}{\sqrt{2\pi \Delta \tilde{\mathcal{S}}_\mathrm{THz}^2(\phi)}}\langle \Psi_\mathrm{THz}|\! :\!\exp\left(-\frac{\left(\mathcal{S}-\hat{\tilde{\mathcal{S}}}_\mathrm{THz}(\phi)\right)^2}{2\Delta \tilde{\mathcal{S}}^2_\mathrm{THz}(\phi)}\right)\! :\!|\Psi_\mathrm{THz}\rangle,
\end{split}\label{probTHz}
\end{align}
where
\begin{align}
\begin{split}
\Delta \tilde{\mathcal{S}}^2_\mathrm{THz}(\phi)&=\langle 0_{\mathrm{THz}}|\hat{\tilde{\mathcal{S}}}_{\mathrm{THz}}^2(\phi)| 0_{\mathrm{THz}}\rangle\\
&=\left[\hat{\alpha}_{\mathrm{THz}}(\phi),\hat{\alpha}_{\mathrm{THz}}^\dagger(\phi)\right]+\left[\hat{\beta}_{\mathrm{THz}}(\phi),\hat{\beta}_{\mathrm{THz}}^\dagger(\phi)\right]+\left(\left[\hat{\alpha}_{\mathrm{THz}}(\phi),\hat{\beta}_{\mathrm{THz}}^\dagger(\phi)\right]+H.c.\right)\label{THzvac}
\end{split}
\end{align}
is given by the vacuum fluctuations of the signal $\hat{\tilde{\mathcal{S}}}_{\mathrm{THz}}(\phi)$. Note that $N+\kappa(\phi)=\tilde{\kappa}(\phi)+\Delta \tilde{\mathcal{S}}^2_\mathrm{THz}(\phi)$. We want to stress that the deconvolution described here does not rely on any assumption about the THz state $|\Psi_\mathrm{THz}\rangle$ other than its initial separability from the NIR state. This deconvolution procedure is therefore also possible for non-Gaussian THz states.

Here, it is apparent that we have chosen $\tilde{\kappa}(\phi)$ such that $P_{\mathrm{THz}}(\mathcal{S},\phi)$ describes the statistics of the operator $\hat{\tilde{\mathcal{S}}}_{\mathrm{THz}}^2(\phi)$, i.e.,
\begin{align}
\langle \Psi_{\mathrm{THz}}|\hat{\tilde{\mathcal{S}}}_{\mathrm{THz}}^k(\phi)| \Psi_{\mathrm{THz}}\rangle=\int_{-\infty}^\infty\!\!\mathrm{d}\mathcal{S}~\mathcal{S}^kP_{\mathrm{THz}}(\mathcal{S},\phi).
\end{align}
Equation~\eqref{conv} together with equation \eqref{prob'} show that the probability distribution $P_{\mathrm{THz}}(\mathcal{S},\phi)$ can be obtained experimentally by deconvolving the measured probability distribution $P(\mathcal{S},\phi)$ of the electro-optic signal with $P'(\mathcal{S},\phi)$, where $P'(\mathcal{S},\phi)$ has to be calculated theoretically. Note that this deconvolution is possible in this case since $P'(\mathcal{S},\phi)$ is Gaussian. We have thus established that, after the just described post-processing of the measured statistics, EOS can be used to obtain the statistics of the THz signal $\hat{\tilde{\mathcal{S}}}_{\mathrm{THz}}(\phi)$. As shown in equation \eqref{transfS}, this THz signal consists of the sum of the two ultrabroadband frequency modes $\alpha(\Omega)$ and $\beta(\Omega)$,
\begin{subequations}
\begin{align}
\hat{\alpha}_{\mathrm{THz}}(\phi)&=\sqrt{N}e^{i\phi}
\int_0^{\Omega_{\mathrm{max}}}\!\!\mathrm{d}\Omega ~\alpha(\Omega)\hat{a}_s(\Omega),\\
\hat{\beta}_{\mathrm{THz}}(\phi)&=\sqrt{N} e^{-i\phi}
\int_0^{\Omega_{\mathrm{max}}}\!\!\mathrm{d}\Omega~ \beta(\Omega)\hat{a}_s(\Omega),
\end{align}
\end{subequations}
with $\alpha(\Omega)=\int_0^\infty\mathrm{d}\omega~ h(\omega)p(\omega,\Omega)$ and $\beta(\Omega)=\int_0^\infty\mathrm{d}\omega~ h^*(\omega)q^*(\omega,\Omega)$. Note that both $\alpha(\Omega)$ and $\beta(\Omega)$ are essentially given by the frequency mode $h(\omega)$, which is shifted down into the THz frequency range through the functions $p(\omega,\Omega)$ and $q^*(\omega,\Omega)$, determined by the Bogoliubov transformation in equation \eqref{btrans}. This results in ultrabroadband frequency modes, whose corresponding temporal modes consist of short (on the order of femtoseconds) few-cycle pulses.

The phase shift $\phi$ controlled by the wave plate in Fig.~\ref{EOS} induces a phase factor 
$e^{i\phi}$ for $\hat{\alpha}_{\mathrm{THz}}(\phi)$ and
$e^{-i\phi}$ for $\hat{\beta}_{\mathrm{THz}}(\phi)$. Therefore, the two respective frequency modes cannot simply be combined into a single mode $\alpha(\Omega)+\beta(\Omega)$ since the phase can not be controlled in this case. We rather need to additionally suppress one of the frequency modes such that $\hat{\tilde{\mathcal{S}}}_{\mathrm{THz}}(\phi)$ can be considered to
consist of $\hat{\alpha}_{\mathrm{THz}}(\phi)$ or $\hat{\beta}_{\mathrm{THz}}(\phi)$ alone. This can be achieved by specific spectral filtering functions $\mathcal{F}(\omega)$ in equation \eqref{signal2}.

Having shown theoretically how the statistics of ultrashort THz states can be generally measured using EOS, we will now provide a particular illustrating example
based on
the Bogoliubov transformation given in Methods~\ref{Methods:Generation} and show how spectral filtering
allows to
to select a specific mode for $\hat{\tilde{\mathcal{S}}}_{\mathrm{THz}}(\phi)$. This Bogoliubov transformation can be utilized here since the squeezing process also relies on the electro-optic effect.

In this case, the ultrabroadband modes $\alpha(\Omega)$ and $\beta(\Omega)$ can be expressed as
\begin{subequations}
\begin{align}
\alpha(\Omega) &= \frac{\sqrt{|\Omega|}}{2\pi}\int_{\Omega_\mathrm{max}}^{\infty}\!\!\mathrm{d}\omega \int_{-\infty}^\infty\!\!\mathrm{d}t \frac{h(\omega)}{\sqrt{\omega}}e^{i\omega \tau^{-1}(t)-i\Omega t},\label{alpha2}\\
\beta(\Omega) &= -\frac{\sqrt{|\Omega|}}{2\pi}\int_{\Omega_{\mathrm{max}}}^{\infty}\!\!\mathrm{d}\omega \int_{-\infty}^\infty\!\!\mathrm{d}t \frac{h^*(\omega)}{\sqrt{\omega}}e^{-i\omega \tau^{-1}(t)-i\Omega t},\label{beta2}
\end{align}
\end{subequations}
where we select $\Omega_\mathrm{max}/(2\pi)=130$~THz and an 
effective squeezing strength $r_\mathrm{eff}=1$ in the conformal time.

\begin{figure}[!t]
\includegraphics[width=\textwidth]{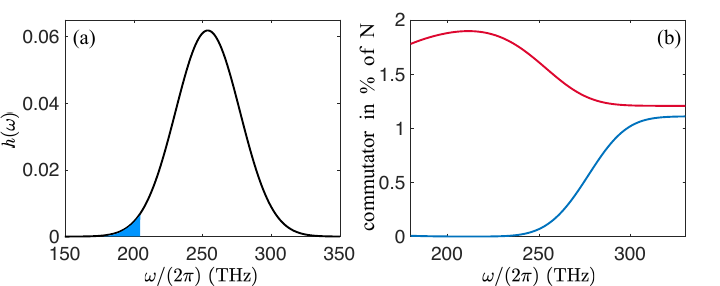}
\caption{Ultrabroadband frequency mode $h(\omega)$ and spectral filtering. (a) The black line shows the unfiltered [$\omega_\mathrm{max}\to \infty$ in equation \eqref{filter}] ultrabroadband mode $h(\omega)$ given by a Gaussian probe of central frequency $\omega_c/(2\pi)=255$~THz and temporal duration at FWHM $\Delta t = 16$~fs. The blue area shows the part of the unfiltered mode that is selected for $\omega_\mathrm{max}/(2\pi)=212$~THz in equation \eqref{filter}. In this case, the THz signal $\hat{\tilde{\mathcal{S}}}
_\mathrm{THz}(\theta)$ is only given by the mode $\beta(\Omega)$ [cf. equation \eqref{signalbeta}]. (b) The red (blue) line depicts the commutator $\left[\hat{\beta}_{\mathrm{THz}}(\phi),\hat{\beta}_{\mathrm{THz}}^\dagger(\phi)\right]/N$ $\left(\left[\hat{\alpha}_{\mathrm{THz}}(\phi),\hat{\alpha}_{\mathrm{THz}}^\dagger(\phi)\right]/N\right)$ relative to the shot noise level $N$ for different $\omega_\mathrm{max}$ in equation \eqref{filter}. \label{QTFig1}}
\end{figure}

Here, we will use a Gaussian probe pulse of the form $E_{\mathrm{p},z}(\omega)=iE_0\exp\left(-\left(\frac{\omega-\omega_c}{\Delta\omega}\right)^2\right)$, with central frequency $\omega_c/(2\pi)=255$~THz and $\Delta\omega/(2\pi)=33$~THz, corresponding to a temporal duration at FWHM of $\Delta t =16$~fs. The corresponding ultrabroadband frequency mode $h(\omega)$ for $\mathcal{F}(\omega)=1$ [cf. \eqref{h}] is shown in Fig.~\ref{QTFig1}a. Figure~\ref{QTFig1}b shows the two commutators $\left[\hat{\alpha}_{\mathrm{THz}}(\phi),\hat{\alpha}_{\mathrm{THz}}^\dagger(\phi)\right]$ and $\left[\hat{\beta}_{\mathrm{THz}}(\phi),\hat{\beta}_{\mathrm{THz}}^\dagger(\phi)\right]$ relative to the shot noise for different spectral filtering functions of the form
\begin{align}
\mathcal{F}(\omega)=\mathrm{H}(\omega_\mathrm{max}-\omega),\label{filter}
\end{align}
where $\mathrm{H}(\omega)$ is the Heaviside function. Here, only the lower part of the probe spectrum is
processed
up to a maximal frequency $\omega_\mathrm{max}$. As expected, measuring only the lower part of the probe spectrum suppresses the mode $\alpha(\Omega)$ in comparison to $\beta(\Omega)$ since the former essentially describes the SFG contribution to the electro-optic process, while $\beta(\Omega)$ describes the DFG contribution. Decreasing $\omega_\mathrm{max}$ in equation \eqref{filter} decreases both the measured shot noise of the probe $N$ and the vacuum fluctuations $\Delta \tilde{\mathcal{S}}^2_\mathrm{THz}(\phi)$ in equation \eqref{THzvac} for $\hat{\tilde{\mathcal{S}}}_\mathrm{THz}(\phi)$. As it is visible in Figure~\ref{QTFig1}b, there exists an optimum at $\omega_\mathrm{max}/(2\pi)\approx 212$~THz for the relative contribution $\left[\hat{\beta}_{\mathrm{THz}}(\phi),\hat{\beta}_{\mathrm{THz}}^\dagger(\phi)\right]/N$. The blue area in Fig.~\ref{QTFig1}a denotes the part of the unfiltered ultrabroadband frequency mode $h(\omega)$ that is used to implement the measurement in this case.

\begin{figure}[!t]
\includegraphics[width=\textwidth]{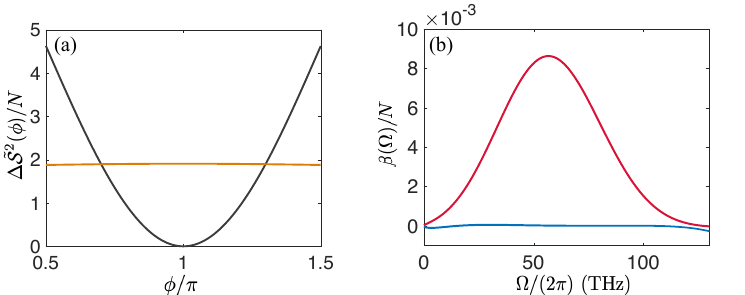}
\caption{Vacuum fluctuations and frequency mode profiles. (a) The black [orange] line depicts the phase dependence of $\Delta \tilde{\mathcal{S}}^2_\mathrm{THz}(\phi)$ for $\omega_\mathrm{max}\to\infty$ $\left[\omega_\mathrm{max}/(2\pi)=212~\mathrm{THz}\right]$. (b) The red (blue) line shows the frequency mode $\beta(\Omega)$ $\left[\alpha(\Omega)\right]$ for the optimal spectral filtering.\label{QTFig2}}
\end{figure}

Figure~\ref{QTFig2}a compares the THz vacuum fluctuations $\Delta \tilde{\mathcal{S}}^2_\mathrm{THz}(\phi)$ relative to the shot noise level $N$ in case of no spectral filtering ($\omega_\mathrm{max}\to\infty$) and according to the optimal spectral filtering with $\omega_\mathrm{max}/(2\pi)=212$~THz. With no spectral filtering applied, $\hat{\tilde{\mathcal{S}}}_\mathrm{THz}(\phi)$ depends on the phase shift $\phi$ because of the interference term $\left(\left[\hat{\alpha}_{\mathrm{THz}}(\phi),\hat{\beta}_{\mathrm{THz}}^\dagger(\phi)\right]+H.c.\right)$ between the two modes in equation \eqref{THzvac}. At the optimal spectral filtering, the mode $\alpha(\Omega)$ can be neglected and the vacuum fluctuations become phase independent as they should be. The frequency modes $\alpha(\Omega)$ and $\beta(\Omega)$ for the optimal spectral filtering are shown in Fig.~\ref{QTFig2}b. The mode $\alpha(\Omega)$ is strongly suppressed in comparison to $\beta(\Omega)$. Therefore, the signal $\hat{\tilde{\mathcal{S}}}_\mathrm{THz}(\phi)$ is essentially given by
\begin{align}
\hat{\tilde{\mathcal{S}}}
_\mathrm{THz}(\phi)\approx\hat{\beta}_\mathrm{THz}(\phi)+\hat{\beta}^\dagger_\mathrm{THz}(\phi)\label{signalbeta}
\end{align}
and the phase shift of this signal can be controlled by $\phi$. Figure~\ref{QTFig2}b also shows that the mode $\beta(\Omega)$ constitutes an ultrabroadband mode with a central frequency of $57$~THz and a FWHM of $54$~THz.

We have shown how EOS can be understood as a homodyne detection of the THz quantum field where the local oscillator is given by an ultrabroadband pulse. The signal in equation \eqref{signalbeta} can measure arbitrary quadratures of the THz quantum field and in particular, we have [cf. equation \eqref{eq:quadratures}]
\begin{align}
\hat{\tilde{\mathcal{S}}}_\mathrm{THz}(\pi)\propto \hat{X}(t_d=0),\\
\hat{\tilde{\mathcal{S}}}_\mathrm{THz}(\pi/2)\propto \hat{P}(t_d=0).
\end{align}
Note that all calculations in this section have been performed for a zero time delay between the THz quantum field and the probe pulse. However, the same arguments hold for arbitrary time delays.

\section{Quantum circuit representation}

\begin{figure}[h]
    \includegraphics[width=16cm]{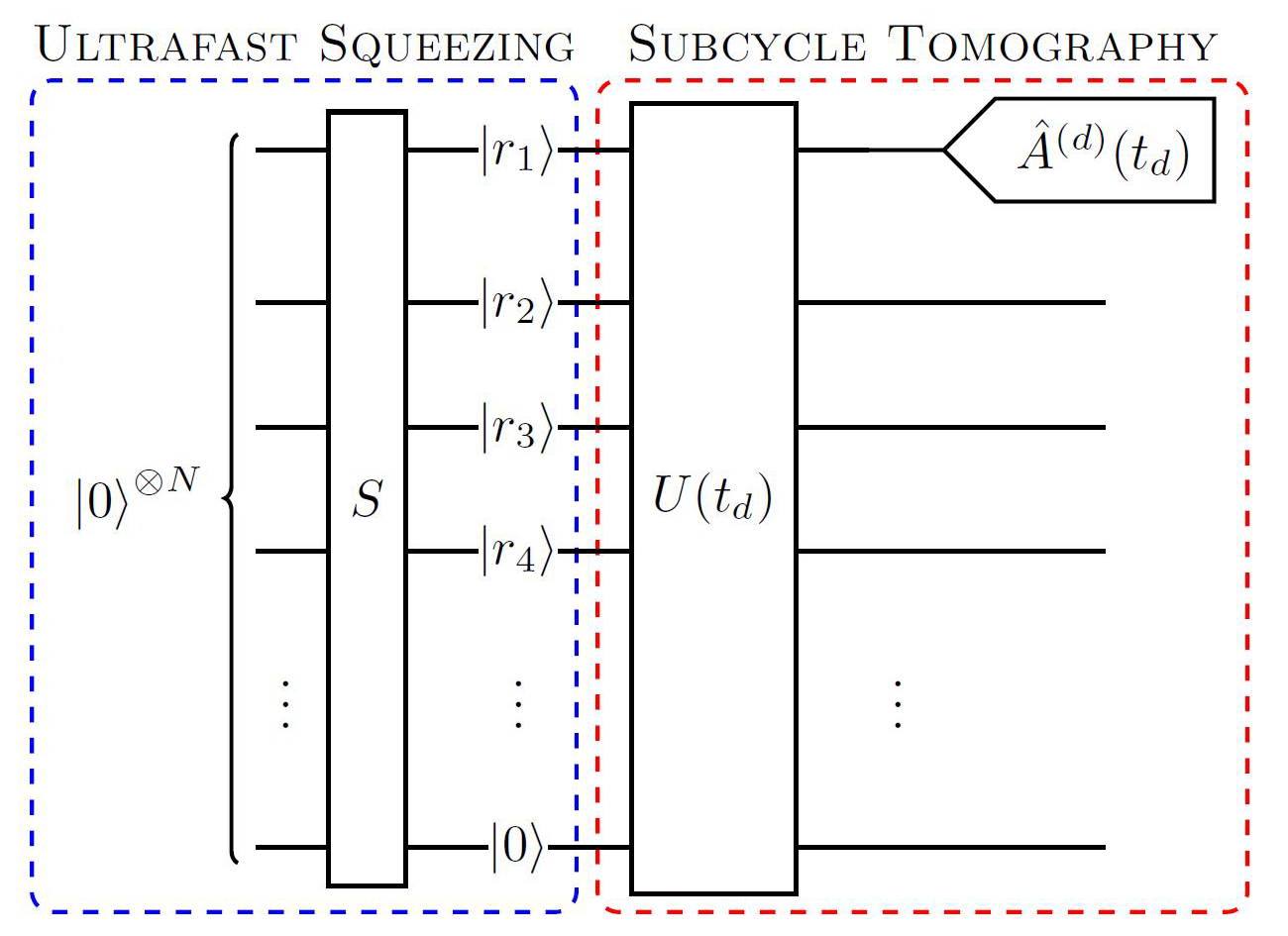}
    \centering
    \caption{\label{Fig:quantum circuit}
    \textbf{Quantum circuit representation of the subcycle tomography of a pulsed squeezed state.} Before the GX in Fig.~\ref{Fig:Subcycle Tomography by BHD}, the field is in the vacuum state, which can be seen as a multimode vacuum state when decomposed into the frequency modes of the relevant frequency range. Ultrafast squeezing operator $S$, resulting physically from the $\chi^{(2)}$ interaction in the GX, transforms the vacuum to a product state of the single-mode squeezed states with squeezing parameters $r_j,\ j=1,2,3,4,\ldots$ \ . A unitary operator $U(t_d)$ depending on the duration and time delay of the probe pulse projects these modes to the detection mode $\hat{A}^{(d)}(t_d)$.}
\end{figure}	

Subcycle tomography of a pulsed squeezed vacuum state, as depicted in Fig.~\ref{Fig:Subcycle Tomography by BHD}, can be represented by a quantum circuit shown in Fig.~\ref{Fig:quantum circuit}. As implied by equation \eqref{eq:BMreduction}, the ultrafast squeezing operator $S$ transforms the multimode vacuum to a separable state, where a few principal modes $j$ are squeezed independently: $S\bigotimes\limits_{k=1}^N\ket{0}_k\approx \bigotimes\limits_{j=1}^l\ket{r_j}\bigotimes\limits_{k=l+1}^N\ket{0}_k$. Here $l$ denotes the number of the principal modes which are required to describe the analyzed quantum state to a desired degree of accuracy. The subsequent subcycle tomography protocol can be described as a single-mode measurement. This measurement corresponds to a projection of the processed signal to the mode $\hat{A}^{(d)}(t_d)$, whereas the mode extraction is realized by the action of the unitary operator $U(t_d)$. Since such a unitary operator represents a multimode beam splitter, $T_j$ in equation \eqref{eq:modaldecompositionofA} can be seen as transmission coefficients.

\begin{figure}[h]
    \includegraphics[width=18cm]{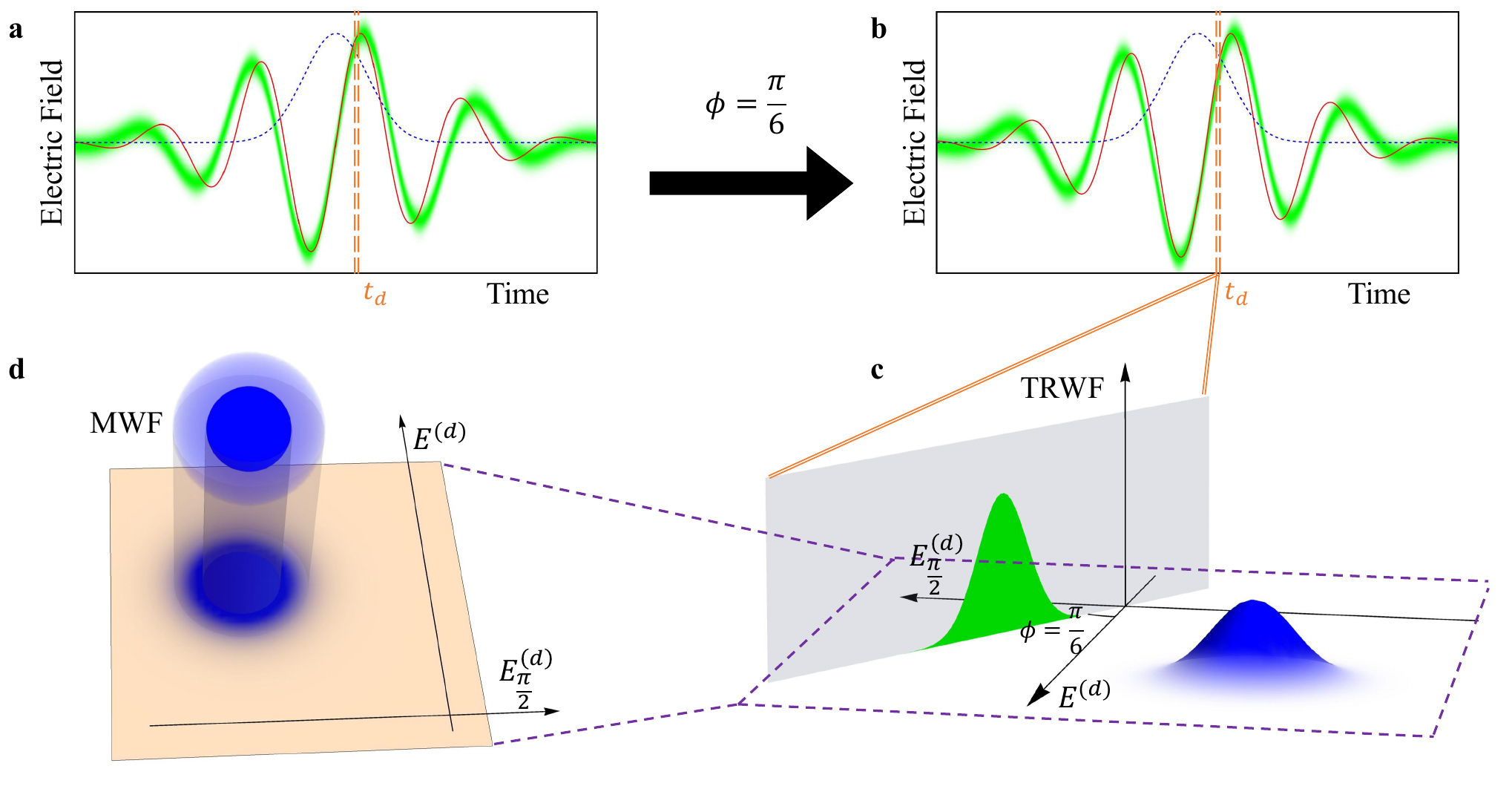}
    \centering
    \caption{\label{Fig:principle of the subcylce tomography}
    \textbf{Quantum statistics of the field and its relation to TRWF, marginal distributions and multimode Wigner function (MWF).} \textbf{a,} Noise trace (green) of a weak coherent few-cycle pulsed electric field (red) as resolved by a probe pulse (dotted blue) with a duration slightly exceeding the half of the optical cycle. \textbf{b,} Noise trace (green) of the $\frac{\pi}{6}$-CEP-shifted electric field (red). \textbf{c,} Quantum statistics of the detected CEP-shifted field corresponds to a  one-dimensional marginal distribution of the TRWF. \textbf{d,} The TRWF constitutes a two-dimensional marginal distribution of the MWF with respect to the plane spanned by vectors representing quadratures $\hat{E}^{(d)}$ and $\hat{E}_{\frac{\pi}{2}}^{(d)}$.}
\end{figure}		

\section{Inverse Radon transform}
The inverse Radon transform has been used to reconstruct the Wigner function from acquired quantum statistic of single-mode quadratures, based on homodyne detection with a variable phase shift $\phi$ of the LO.
%
%
%
The projection of the Wigner function
onto the axis with polar angle $\phi$ equals to the probability distribution of the $\phi$-phase-rotated quadrature, which is measured.  The projection-slice theorem states that the slice of the characteristic function of the Wigner function along
that axis
is the Fourier transform of the probability distribution of the $\phi$-phase-rotated quadrature, so that the  inverse Radon transform can be used for the reconstruction. In the case of subcycle tomography, the projection of the TRWF can be assumed \cite{LeonhardtEssential} to be the probability distribution of $\hat{X}_\phi(t_d)$, as depicted in Figs.~\ref{Fig:principle of the subcylce tomography}b,c: $[P_\phi W](q)=\mathrm{pr}_{\phi}(q)$, where $W$ represents the TRWF, $P_\phi$ denotes projection onto the axis with angle $\phi$, and $\mathrm{pr}_{\phi}$ is the probability distribution of $\hat{X}_\phi(t_d)$. The projection-slice theorem means $F_1P_\phi=S_\phi F_2$, where $F_1$ ($F_2$) denotes the one(two)-dimensional Fourier transform and $S_\phi$ refers to the slice along the axis with angle $\phi$. The characteristic function of the TRWF on the axis with angle $\phi$ equals then to the Fourier transform of the probability distribution of $\hat{X}_\phi(t_d)$, as follows from
\begin{equation}\label{eq:Inverse radon of the TRWF}
\widetilde{W}(\xi \cos \phi, \xi \sin \phi;t_d)=[S_\phi F_2 W](\xi)=[F_1 P_\phi W](\xi)=\widetilde{\mathrm{pr}}_{\phi}(\xi)
\end{equation}
with $\widetilde{\mathrm{pr}}_{\phi}=F_1 \mathrm{pr}_{\phi}$. Therefore, acquiring probability distributions of $\hat{X}_\phi(t_d)$ for all $\phi$, we can reconstruct $W(x,p;t_d)$ from its characteristic function.

$\widetilde{W}(u,v;t_d)$ reconstructed from equation \eqref{eq:Inverse radon of the TRWF} can be seen as equation \eqref{eq:characteristicfunction} in the Hilbert space of the ultrabroadband quantum light, which can be effectively represented as $\mathcal{H}_d\otimes\bigotimes_{j=2}^N\mathcal{H}_j$ corresponding to the output of the circuit shown in the Fig.~\ref{Fig:quantum circuit}, where $\mathcal{H}_d$ is the Hilbert space of the detection mode and $\mathcal{H}_j$ is the Hilbert space of $j$th mode. The Born rule of $\hat{X}_\phi(t_d)$ measurement reads $\mathrm{pr}_{\phi}(q)=\sum\limits_{k}\bra{q,\phi,k;t_d} \hat{\rho} \ket{q,\phi,k;t_d}$ with degenerate eigenstates $\ket{q,\phi,k;t_d}$ defined by $\hat{X}_\phi(t_d)\ket{q,\phi,k;t_d}=q\ket{q,\phi,k;t_d}$. Such a state can be represented as a product state $\ket{q,\phi,k;t_d}=\ket{q,\phi;t_d}\otimes\ket{k}$, where $\hat{X}_\phi(t_d)\ket{q,\phi;t_d}=q\ket{q,\phi;t_d}$ with $\ket{q,\phi;t_d} \in \mathcal{H}_d$ and  $\ket{k} \in \otimes_{j=2}^N\mathcal{H}_j$. Here $\bra{k}\ket{k'}=\delta_{kk'}$ and $\sum \limits_k \ket{k}\bra{k}=\mathds{1}_\mathrm{rest}$, where $\mathds{1}_\mathrm{rest}$ is the identity operator in $\otimes_{j=2}^N\mathcal{H}_j$. Equation \eqref{eq:Inverse radon of the TRWF} then gives
\begin{equation}\label{eq:Slice relation of the TRWF}
\begin{split}
\widetilde{\mathrm{pr}}_{\phi}(\xi)&=\sum\limits_k\int \mathrm{d}q \bra{q,\phi,k;t_d} \hat{\rho} \ket{q,\phi,k;t_d}e^{-i\xi q}\\
&=\sum\limits_k\int \mathrm{d}q \bra{q,\phi,k;t_d} \hat{\rho} e^{-i\xi [\cos\phi \hat{X}(t_d)+\sin\phi \hat{P}(t_d)]} \ket{q,\phi,k;t_d}\\
&=\sum\limits_k \bra{k} \int \mathrm{d}q \bra{q,\phi;t_d}
 \hat{\rho}  e^{-i\xi \left[\cos\phi \hat{X}(t_d)+\sin\phi \hat{P}(t_d)\right]} \ket{q,\phi;t_d} \ket{k}\\
 &= \mathrm{Tr}_\mathrm{rest} \left\{ \int \mathrm{d}q \bra{q,\phi;t_d}
 \hat{\rho}  e^{-i\xi \left[\cos\phi \hat{X}(t_d)+\sin\phi \hat{P}(t_d)\right]} \ket{q,\phi;t_d} \right\}\\
 &=\mathrm{Tr}_\mathrm{rest}\left\{ \mathrm{Tr}_{d} \left\{ \hat{\rho}  e^{-i\xi \left[\cos\phi \hat{X}(t_d)+\sin\phi \hat{P}(t_d)\right]} \right\} \right\}\\
 &= \mathrm{Tr} \left\{ \hat{\rho}  e^{-i\xi \left[\cos\phi \hat{X}(t_d)+\sin\phi \hat{P}(t_d)\right]} \right\}=\widetilde{W}(\xi \cos \phi, \xi \sin \phi;t_d),
\end{split}
\end{equation}
where $\mathrm{Tr}_\mathrm{rest}$ ($\mathrm{Tr}_d$) means partial trace over $\otimes_{j=2}^N\mathcal{H}_j$ ($\mathcal{H}_d$). Completeness and orthogonality of $\{\ket{q,\phi;t_d}\mid \forall q \in \mathbb{R} \}$ in $\mathcal{H}_d$ ensure $\int \mathrm{d}q \bra{q,\phi;t_d} \hat{A} \ket{q,\phi;t_d}=\mathrm{Tr}_d\big[\hat{A}\big]$, where $\hat{A}$ is any operator acting in the total Hilbert space. Such properties of $\ket{q,\phi;t_d}$ can be proved from the number-state representation (cf. ref.~\cite{Vogel2006}, p.~104), which can be shown by a recurrence relation $\braket{n+1}{q,\phi;t_d}=\bra{n}\hat{A}^{(d)\dagger}(t_d)\ket{q,\phi;t_d}$ with the initial condition $\braket{0}{q,\phi;t_d}$ and $\ket{n}$ being the number states in $\mathcal{H}_d$. Proceeding in an analogous way as when obtaining the vacuum wave function for a simple harmonic oscillator, one can show that the solution of $\bra{q,\phi;t_d}\hat{A}^{(d)}(t_d)\ket{0}=0$ reads $\braket{q,\phi;t_d}{0}=\pi^{-\frac{1}{4}}e^{-\frac{1}{2}q^2}$. It follows from $\bra{q,\phi;t_d}\hat{X}_{\phi+\frac{\pi}{2}}(t_d)\ket{\psi}=-i\frac{\partial}{\partial q}\braket{q,\phi;t_d}{\psi}$ for any $\ket{\psi}\in \mathcal{H}_d$, which is a consequence of the commutation property $[\hat{X}_{\phi}(t_d),\hat{X}_{\phi+\frac{\pi}{2}}(t_d)]=i$. Recurrence relation
\begin{equation}\label{eq:recurrence relation of the quadrature eigenstate}
\bra{q,\phi;t_d}\hat{A}^{(d)\dagger}(t_d)\ket{n}=\sqrt{n+1}\braket{q,\phi;t_d}{n+1}=\frac{1}{\sqrt{2}}\left(e^{-i\phi}q-e^{-i\phi}\frac{\partial}{\partial q}\right) \braket{q,\phi;t_d}{n}
\end{equation}
with the initial condition $\braket{q,\phi;t_d}{0}=\pi^{-\frac{1}{4}}e^{-\frac{1}{2}q^2}$ ensures $\braket{q,\phi;t_d}{n}=\frac{H_n(q)}{\sqrt{2^n n! \sqrt{\pi}}}e^{-\frac{q^2}{2}}e^{-i n\phi}=\psi_n(q)e^{-i n\phi}$, where $\psi_n(q)$ is the \textit{n}th-order Hermite function. The number-state representation of the quadrature eigenstates in $\mathcal{H}_d$, i.e. $\ket{q,\phi;t_d}=\sum\limits_{n=0}\psi_n(q) e^{i n\phi}\ket{n}$, leads to their orthogonality and completeness, which follow directly from such properties of $\psi_n(q)$. Therefore, the function reconstructed by the inverse Radon transform represents equation \eqref{eq:characteristicfunction}.

\section{Relation of the TRWF to the multimode Wigner function}


The subcycle measurements capture the CEP-shifted electric field of a few-cycle coherent light properly even if the probe pulse duration is comparable with the length of the optical cycle, as seen in Figs.~\ref{Fig:principle of the subcylce tomography}a,b. The field probability distributions at any time moment $t_d$ represent projections of the TRWF,
as illustrated in Figs.~\ref{Fig:principle of the subcylce tomography}b,c and implied by equation \eqref{eq:Inverse radon of the TRWF}.
The TRWF, in its turn, is a projection of the general multimode Wigner function (MWF) to the plane spanned by the vectors corresponding to the detected CEP-shifted electric fields, as shown in Fig.~\ref{Fig:principle of the subcylce tomography}d. Thus, the TRWF constitutes a two-dimensional marginal distribution of the MWF.
The MWF for a $N$ mode state $\hat{\rho}_N$ can be defined from the characteristic function \cite{WeedbrookGaussian}
\begin{eqnarray}\label{eq:characteristicfunctionofMWF}
\widetilde{W}_N(\mathbf{u},\mathbf{v})=\text{tr}\Big[\hat{\rho}_N\text{exp}\big\{-i\;\hat{\mathbf{x}}\mathbf{u}^\top -i\;\hat{\mathbf{p}}\mathbf{v}^\top \big\}\Big],
\end{eqnarray}
where $\mathbf{u}=(u_1,u_2,\cdots,u_N)$ and $\mathbf{v}=(v_1,v_2,\cdots,v_N)$ are vectors of the Fourier phase-space coordinates.
Further, $\hat{\mathbf{x}}=(\hat{x}_1,\hat{x}_2,\cdots,\hat{x}_N)$ and $\hat{\mathbf{p}}=(\hat{p}_1,\hat{p}_2,\cdots,\hat{p}_N)$ represent quadrature vectors. The \textit{n}th component of each vector is the respective quadrature operator of the \textit{n}th frequency mode.	Proceeding in a similar way as in equation \eqref{eq:Slice relation of the TRWF}, we can see that the Fourier-transformed probability distribution of $\hat{\mathbf{r}}=\sum\limits_{k=1}^N(r_k\,\hat{x}_k+r_{N+k}\,\hat{p}_k)$ gives $\widetilde{W}_N(\mathbf{u},\mathbf{v})$ taken at the straight line connecting the origin and vector $\mathbf{r}=(r_1,r_2,\cdots,r_{2N})$.
Therefore, the Fourier-transformed probability distributions of $\hat{E}^{(d)}_\phi(t_d)$ for all $\phi$ gives $\widetilde{W}_N(\mathbf{u},\mathbf{v})$ at the plane spanned by
\begin{eqnarray}\label{eq:vectorofE}
\begin{split}
\mathbf{E}^{(d)}(t_d)=\sqrt{2C}\delta\omega & \big(\sqrt{\delta\omega}  \widetilde{R}(\delta\omega) \,\text{sin}(\delta\omega \,t_d),\cdots,\sqrt{N\delta\omega} \widetilde{R}(N\delta\omega)\,\text{sin}(N\delta\omega \,t_d),\\&-\sqrt{\delta\omega} \widetilde{R}(\delta\omega)\,\text{cos}(\delta\omega \,t_d),\cdots,-\sqrt{N\delta\omega} \widetilde{R}(N\delta\omega)\,\text{cos}(N\delta\omega \,t_d)\big)
\end{split}
\end{eqnarray}
and
\begin{eqnarray}\label{eq:vectorofEshift}
\begin{split}
\mathbf{E}^{(d)}_{\frac{\pi}{2}}(t_d)=\sqrt{2C}\delta\omega  & \big(-\sqrt{\delta\omega} \widetilde{R}(\delta\omega)  \,\text{cos}(\delta\omega \,t_d),\cdots,-\sqrt{N\delta\omega} \widetilde{R}(N\delta\omega)\,\text{cos}(N\delta\omega \,t_d),\\&-\sqrt{\delta\omega} \widetilde{R}(\delta\omega)\,\text{sin}(\delta\omega \,t_d),\cdots,-\sqrt{N\delta\omega} \widetilde{R}(N\delta\omega)\,\text{sin}(N\delta\omega \,t_d)\big),
\end{split}
\end{eqnarray}
where the mode spacing $\delta \omega$ satisfies $1\gg\delta \omega \,\delta_d \gg \frac{1}{N}$, the constant $C$ was introduced in Section \ref{Methods:Sampling} of Methods and the gating $R(t)$ is assumed to be time-symmetric so that $\widetilde{R}(\omega)=\widetilde{R}(-\omega)$, with
$\widetilde{R}(\omega)=\int_{-\infty}^{\infty}\mathrm{d}t \, R(t)e^{i\omega t}$.
Since slices of $\widetilde{W}(u,v;t_d)$ also represent Fourier-transformed probability distributions of $\hat{E}^{(d)}_\phi(t_d)$ with the normalization as in equation \eqref{eq:quadratures} and decomposition coefficients of $-\mathcal{N}[u\hat{E}^{(d)}(t_d)+v\hat{E}^{(d)}_{\frac{\pi}{2}}(t_d)]$ with respect to $\hat{\mathbf{x}}$ and $\hat{\mathbf{p}}$ are
\begin{eqnarray}
\begin{split}
\mathbf{u}_d=\frac{\delta\omega}{\sqrt{\int_0^\infty\mathrm{d}\omega\,\omega\lvert \widetilde{R}(\omega)\rvert^2}} & \big(\sqrt{\delta\omega} \widetilde{R}(\delta\omega)(u\,\text{cos}(\delta\omega \,t_d)-v\,\text{sin}(\delta\omega \,t_d)) ,\\& \cdots ,  \sqrt{N\delta\omega} \widetilde{R}(N\delta\omega)(u\,\text{cos}(N\delta\omega \,t_d)-v\,\text{sin}(N\delta\omega \,t_d))\big),
\end{split}
\end{eqnarray}
\begin{eqnarray}
\begin{split}
\mathbf{v}_{d}=\frac{\delta\omega}{\sqrt{\int_0^\infty\mathrm{d}\omega\,\omega\lvert \widetilde{R}(\omega)\rvert^2}} & \big(\sqrt{\delta\omega} \widetilde{R}(\delta\omega)(u\,\text{sin}(\delta\omega \,t_d)+v\,\text{cos}(\delta\omega \,t_d)),\\& \cdots, \sqrt{N\delta\omega} \widetilde{R}(N\delta\omega)(u\,\text{sin}(N\delta\omega \,t_d)+v\,\text{cos}(N\delta\omega \,t_d))\big),
\end{split}
\end{eqnarray}
equation \eqref{eq:characteristicfunctionofMWF} gives $\widetilde{W}_N(\mathbf{u}_d,\mathbf{v}_d)=\widetilde{W}(u,v;t_d)$. Figures \ref{Fig:principle of the subcylce tomography}c,d illustrate the Fourier transform of this relation, which means that the TRWF can be seen as a two-dimensional marginal distribution of the MWF. As $t_d$ varies, the plane at which $\widetilde{W}_N(\mathbf{u},\mathbf{v})$ reveals $\widetilde{W}(u,v;t_d)$ rotates.

\section{Field modes
corresponding to
the principal modes}

\begin{figure}[h]\includegraphics[width=16cm]{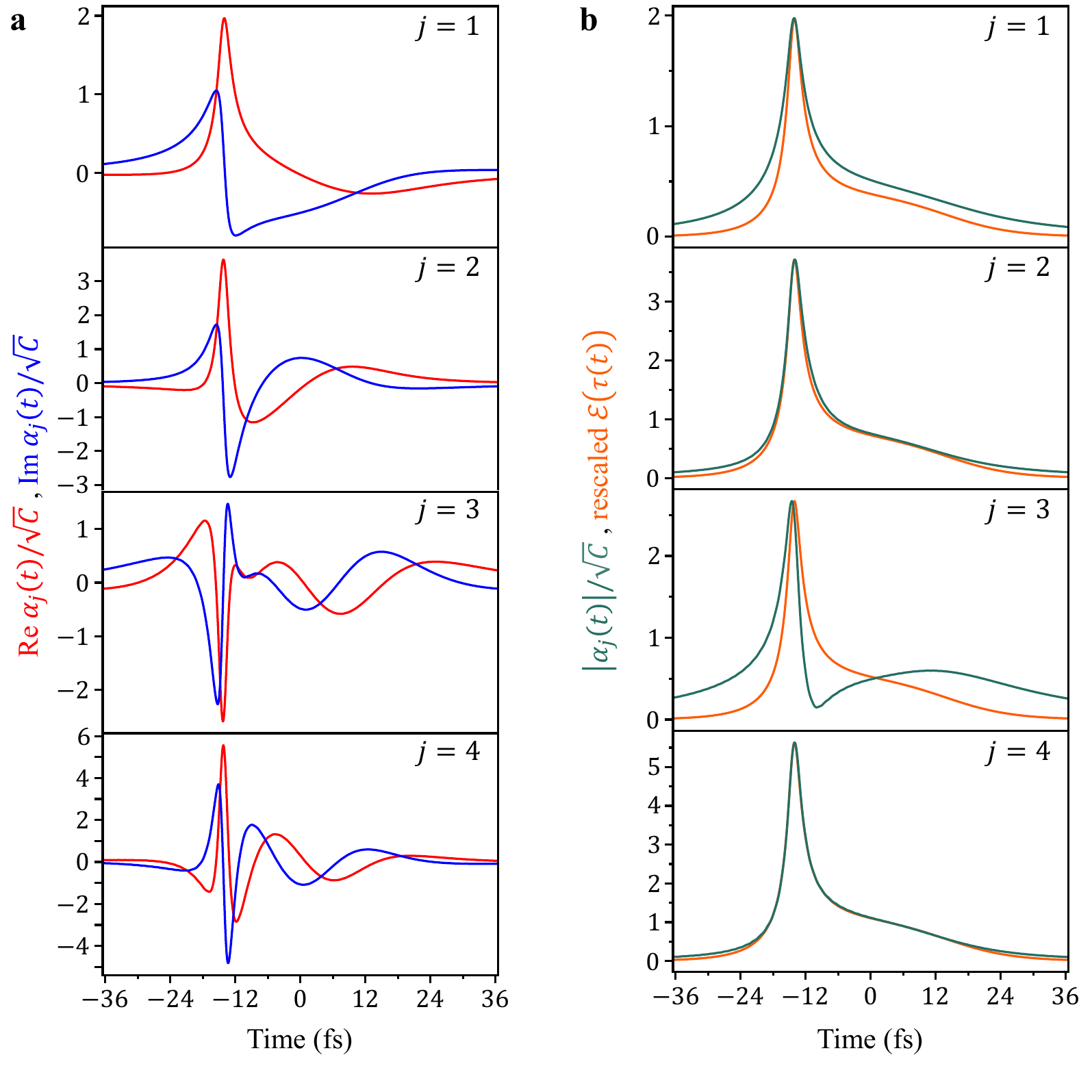}
	\centering
	\caption{\label{Fig:alphaj conformal driving field}
		\textbf{Temporally localized field modes corresponding to the dominant principal modes.}
  \textbf{a,} Real (red) and imaginary (blue) part of $\alpha_j(t)$. \textbf{b,} Absolute value of $\alpha_j(t)$ (green) and $\mathcal{E}(\tau(t))$ rescaled to have the same maximum value (orange).
  The constant $C$ used for the normalization is introduced in Section \ref{Methods:Sampling} of Methods. Parameters: $r_\mathrm{eff}=5$ and $\delta_d=16\,$fs (cf. Section \ref{Methods:Generation} of Methods)
  }
\end{figure}	

The pulsed squeezed state can be decomposed into single-mode squeezed states of the principal modes, which represent output modes of the Bloch-Messiah reduction. To comprehend the dynamics observed by the subcycle tomography, the field modes $\alpha_j(t)=[\hat{E}^{(+)}(t),\hat{b}^\dagger_j]\,$ are computed for four dominant principal modes in Fig.~\ref{Fig:alphaj conformal driving field}a.
More oscillations can be observed for higher-order modes, especially pronounced
near $t_d=-14\,$fs. Since the coefficients $\theta_j$ are proportional to the cross-correlation of $R(t)$ and $\alpha_j(t)$ as shown in equation \eqref{eq:transmittance}, which is the same as the convolution here since $R(t)$ is even, such oscillations of the third and fourth mode get washed out for $\delta_p=5.8\,$fs whereas the slower oscillations of the first and second mode can still be resolved (cf. Fig.~\ref{Fig:TRWF_dynamics}a and Fig.~\ref{Fig:alphaj conformal driving field}a). 
In result, the
transmission
of the first and second mode
dominates the detected signal in the time interval around $t_d=-14\,$fs.
In the case of $\delta_p$ shorter than the scale of oscillations in both the real and imaginary parts of $\alpha_j(t)$,
$T_j(t_d)$ should follow $\lvert \alpha_j(t) \rvert^2$ in Fig.~\ref{Fig:alphaj conformal driving field}b.
Consequently, we find from this figure that $T_1(t_d)$ and $T_2(t_d)$ well resemble the temporal shape of  $\mathcal{E}^2(\tau(t))$.
After $t_d=-10\,$fs, also $T_3(t_d)$ and $T_4(t_d)$ start to follow $\lvert \alpha_{3}(t) \rvert^2$ and $\lvert \alpha_{4}(t) \rvert^2$, respectively, because the oscillation cycles of $\alpha_3(t)$ and $\alpha_4(t)$ increase sufficiently beyond $\delta_p$.
Especially, $T_3(t_d)$ shows a bump near $t_d=15\,$fs and $T_4(t_d)$ prevails over the other transmission coefficients in the interval $t_d=-6\sim12\,$fs
(note the different scale of the vertical axis in the bottom plot of Fig.~\ref{Fig:alphaj conformal driving field}b).

To understand why $\lvert \alpha_1(t)\rvert, \lvert \alpha_2(t)\rvert$ and $\lvert \alpha_4(t)\rvert$ follow $\mathcal{E}(\tau(t))$, it is instructive to analyze the input-output relation for the generated field, $\hat{E}_
\mathrm{out}(t)=\frac{\mathcal{E}(\tau(t))}{\mathcal{E}(t)} \hat{E}_\mathrm{in}(\tau(t))$, provided in Section \ref{Methods:Generation} of Methods, with $\frac{d \tau}{dt}(t)=\frac{\mathcal{E}(\tau(t))}{\mathcal{E}(t)}$\cite{KizmannSubcycle}. The Bloch-Messiah reduction of the field relation leads to a bijection between the input and output field modes for each $j$,
\begin{equation}\label{eq:BM reduction of field modes}
\alpha_{\mathrm{out},j}(t)=\frac{\mathcal{E}(\tau(t))}{\mathcal{E}(t)} \left[\cosh r_j \alpha_{\mathrm{in},j}(\tau(t))-\sinh r_j \alpha^*_{\mathrm{in},j}(\tau(t))\right].
\end{equation}
Since a time symmetric driving, $\mathcal{E}(-t)=\mathcal{E}(t)$, implies $\tau(-t)=-\tau^{-1}(t)$\cite{KizmannSubcycle},  equation \eqref{eq:BM reduction of field modes} is also satisfied for $\alpha_{\mathrm{out},j}(t) \rightarrow i\alpha_{\mathrm{in},j}^*(-t)$ and $\alpha_{\mathrm{in},j}(t) \rightarrow i\alpha_{\mathrm{out},j}^*(-t)$. Then, the sign freedom of the Bloch-Messiah reduction gives $\alpha_{\mathrm{out},j}(t)=\pm i\alpha_{\mathrm{in},j}^*(-t)$. With that, equation \eqref{eq:BM reduction of field modes} becomes a functional equation for $\alpha_{\mathrm{in},j}(t)$, i.e.
\begin{equation}\label{eq:input field mode}
\pm i \alpha^*_{\mathrm{in},j}(-t)=\frac{\mathcal{E}(\tau(t))}{\mathcal{E}(t)} \left[\cosh r_j \alpha_{\mathrm{in},j}(\tau(t))-\sinh r_j \alpha^*_{\mathrm{in},j}(\tau(t))\right].
\end{equation}
In order to simplify equation \eqref{eq:input field mode}, the 0th-order
Taylor series for $\cosh r_j$ and $\sinh r_j$  at $r_j=0$, $\cosh r_j\simeq 1$ and $\sinh r_j\simeq 0$,  can be used. This leads to
\begin{equation}\label{eq:simple input field mode eq}
\pm i \alpha^*_{\mathrm{in},j}(-t)=\frac{\mathcal{E}(\tau(t))}{\mathcal{E}(t)} \alpha_{\mathrm{in},j}(\tau(t)).
\end{equation}
One of the solutions of equation \eqref{eq:simple input field mode eq} reads
\begin{equation}
\alpha_{\mathrm{in},j}(t)=c_j\, \mathcal{E}(\tau^{-1}(t))e^{if_j(t)},\label{eq:solution of input field mode}
\end{equation}
implying then also
\begin{equation}
\alpha_{\mathrm{out},j}(t)= c_j\, \mathcal{E}(\tau(t)) e^{i(\pm \frac{\pi}{2}-f_j(-t))}.\label{eq:solution of output field mode}
\end{equation}
Here constants $c_j$ and phase functions $f_j(t)$ fulfill the condition $\pm \frac{\pi}{2}-f_j(-t)=f_j(\tau(t))$. Such constants $c_j$ and functions $f_j(t)$ should be selected to satisfy the commutation relations for the input (output) modes: $\left[\hat{a}_j,\hat{a}_k\right]=0$ ($[\hat{b}_j,\hat{b}_k]=0$) and $\big[\hat{a}_j,\hat{a}^\dagger_k\big]=\delta_{jk}$ ($\big[\hat{b}_j,\hat{b}^\dagger_k\big]=\delta_{jk}$). We can see that the shapes of $\alpha_1(t), \alpha_2(t)$ and $\alpha_4(t)$ originate basically from
\eqref{eq:solution of output field mode}, whereas minor corrections are caused
by the higher-order terms of the Taylor series for $\cosh r_j$ and $\sinh r_j$, cf.
Fig.~\ref{Fig:alphaj conformal driving field}b. As the solutions of equation \eqref{eq:simple input field mode eq} are not uniquely given by \eqref{eq:solution of output field mode}, it happens that $\alpha_3(t)$ does not possess the shape determined by \eqref{eq:solution of output field mode}.

\section{Dynamics of the ultrafast squeezed state}

\begin{figure}[h]\includegraphics[width=18cm]{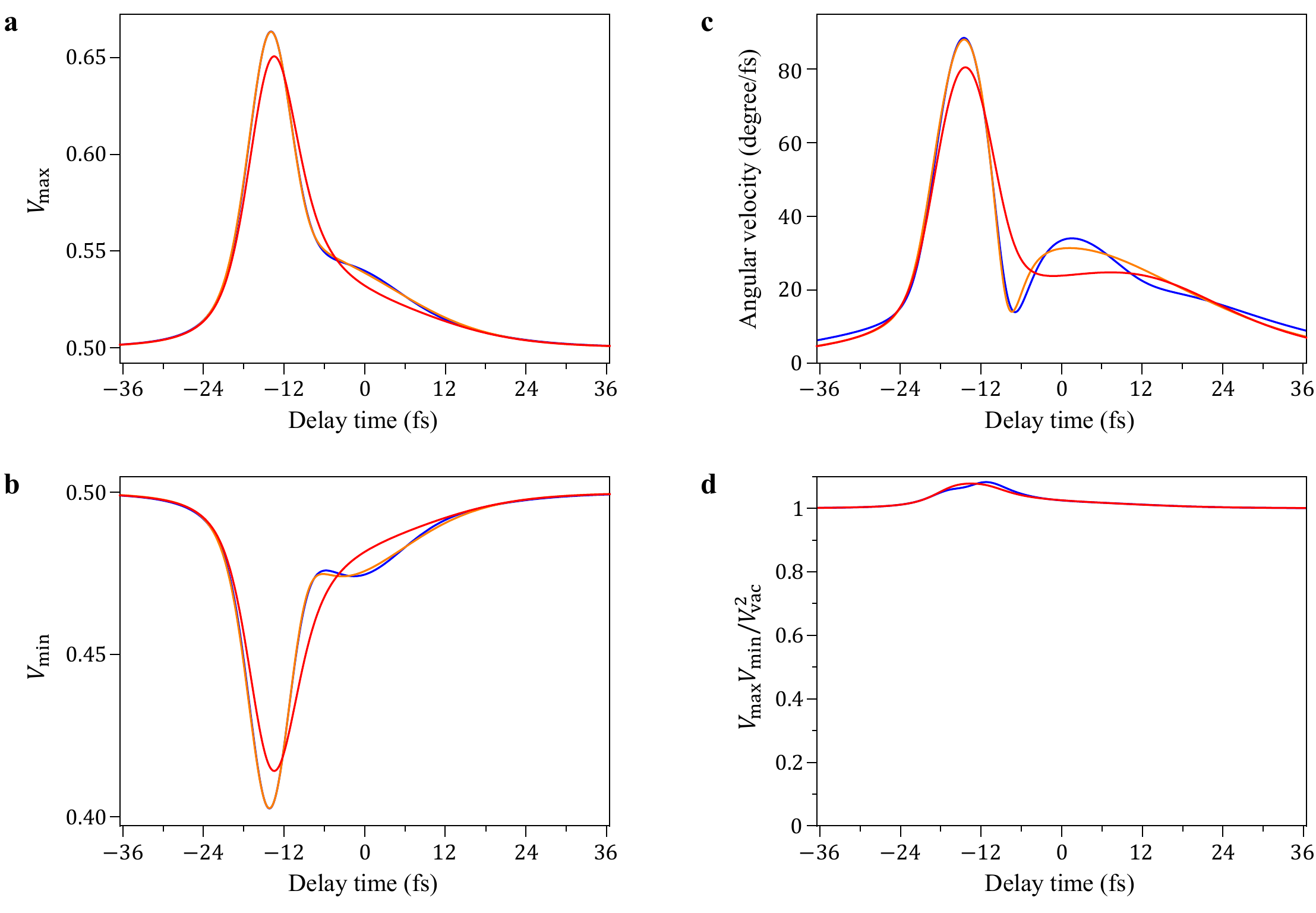}
    \centering
    \caption{\label{Fig:Dynamics of the squeezed state}
    \textbf{Dynamics of the characteristic quantities for the pulsed squeezed light analyzed in terms of the contributions of the dominant modes.} Contributions up to the first (red), second (orange), third (dotted green), and fourth (blue) order are included. Maximum $V_\mathrm{max}$ (\textbf{a}) and minimum $V_\mathrm{min}$ (\textbf{b}) variance of the generalized quadratures.
    \textbf{c,} Angular velocity of the squeezing axis.
    \textbf{d,} Deviation of $V_\mathrm{max}V_\mathrm{min}$ from $V_\mathrm{vac}^2$ showing the increase in uncertainty. Here, only two cases are plotted: the first-order mode (red) and all modes up to the fourth order (blue). The parameter values are as in Fig.~\ref{Fig:TRWF_dynamics}. }
\end{figure}	

$W_\mathrm{psq}(x,p;t_d)$ shown in Fig.~\ref{Fig:TRWF_dynamics}b can be parameterized by the angle of the squeezing axis as well as the minimum and maximum quadrature variances. Since $W_\mathrm{psq}(x,p;t_d)$ is a normal distribution, these three quantities are sufficient to characterize its dynamics. They are computed in Figs.~\ref{Fig:Dynamics of the squeezed state}a,b,c
varying how many of
squeezed states in inferior modes are regarded as vacuum states, to illuminate the contributions of the modes to the dynamics.
Judging by the values of the squeezing parameters $r_j$ in equation \eqref{eq:BMreduction}, the dynamics originates mainly from the first mode, while the correction comes mostly from the second mode. Even if we include contributions from the squeezed states in modes of the order higher than four, there is no perceivable change in resulting dynamics of the axis angle and variances. Thermalization effect, i.e. an appearance of thermal photons in the detection mode, manifests itself in the deviation of $V_\mathrm{max}V_\mathrm{min}$ from $V_\mathrm{vac}^2$. If we relate the obtained TRWF to  the Wigner function of a squeezed thermal state\cite{Kim1989}, this deviation $V_\mathrm{max}V_\mathrm{min}/V_\mathrm{vac}^2-1$ amounts to $4\bar{n}(\bar{n}+1)$, where $\bar{n}$ is the average photon number of the thermal state before squeezing. As shown in Fig.~\ref{Fig:Dynamics of the squeezed state}d, the thermalization effect  is insignificant for the considered pulsed squeezed state.

\section{Single-photon subtraction}
				
\begin{figure}[h]
  \includegraphics[width=16cm]{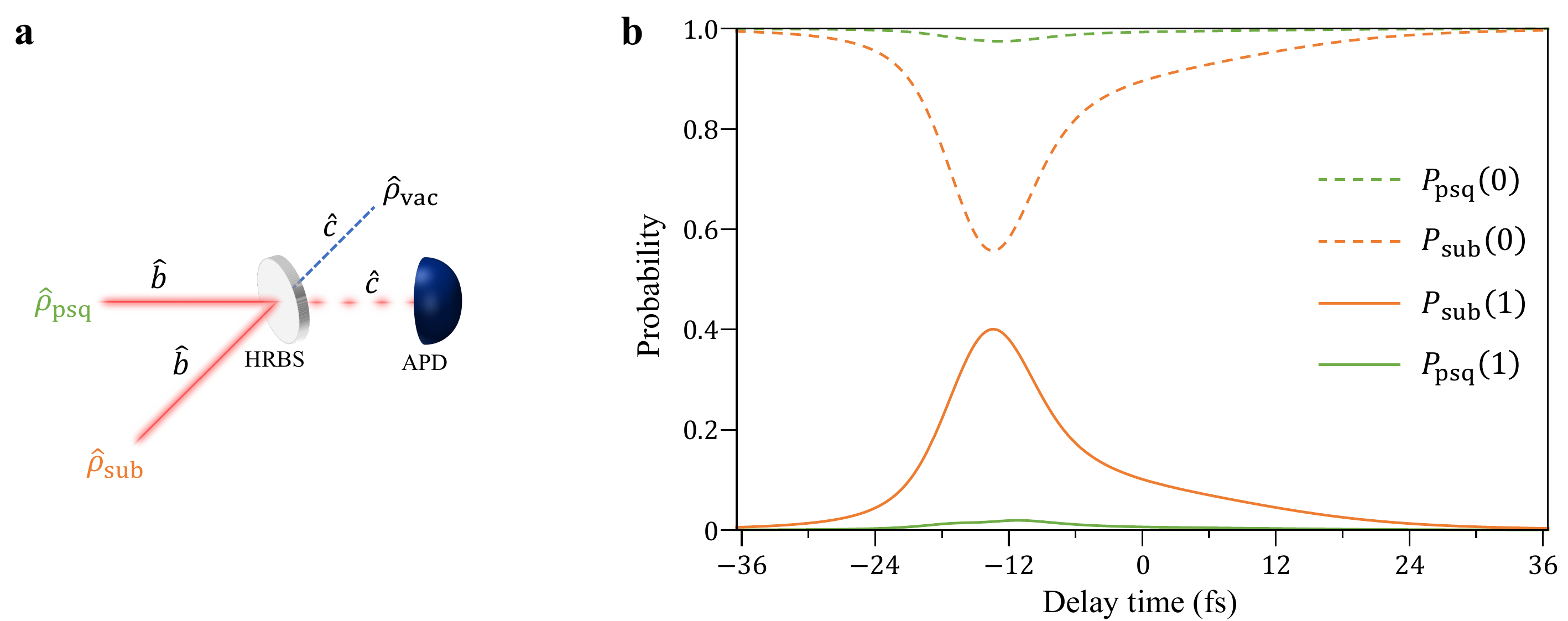}
    \centering
    \caption{\label{Fig:Single photon subtraction}\textbf{Scheme of single-photon subtraction for pulsed squeezed state and photon number dynamics before and after the subtraction.}
    \textbf{a,} Photon subtraction scheme realized by the HRBS and APD. Photon annihilation operators of incident and reflected light (vacuum input and transmitted light) are denoted by $\hat{b}$ ($\hat{c}$).
    \textbf{b,} Probabilities of the photon number measurement corresponding to the vacuum outcome (dashed line) and the single-photon outcome (solid line) for the detection mode $\hat{A}^{(d)}(t_d)$ of the pulsed squeezed state $P_\mathrm{psq}$ (green) and the photon-subtracted state derived from it $P_\mathrm{sub}$ (orange). The parameter values are as in Fig.~\ref{Fig:TRWF_dynamics}.}
\end{figure}
Single-photon subtraction at the right end of Fig.~\ref{Fig:Subcycle Tomography by BHD}a is realized via postselection of the APD signal after the HRBS\cite{Averchenko2016}. A detailed scheme is depicted in Fig.~\ref{Fig:Single photon subtraction}a. 
To simplify the description of the unitary operation implemented by the HRBS, we denote photon annihilation operators of the reflected (transmitted) light by $\hat{b}$ ($\hat{c}$). The HRBS operator can then be approximated by
\begin{eqnarray}
\begin{split}
\hat{B}_\mathrm{HR}&=\text{exp}\Big[\kappa \int_0^\infty \big(\hat{b}^\dagger(\omega)\hat{c}(\omega)-\hat{b}(\omega)\hat{c}^\dagger(\omega)\big) \mathrm{d}\omega\Big]\\
&\approx\mathds{1}+\kappa \sum\limits_j\left(\hat{b}_j^\dagger\hat{c}_j-\hat{b}_j\hat{c}_j^\dagger\right),
\end{split}
\end{eqnarray}
where $\hat{c}_j=\int_0^\infty\mathrm{d}\omega\,\psi_j(\omega)\hat{c}(\omega)$ represent the same temporal modes as $\hat{b}_j$ and $\kappa=\arccos R \ll 1$ for a high reflection coefficient $R\approx1$. After the HRBS, the light state becomes $\hat{\rho}'=\hat{B}_\mathrm{HR} (\hat{\rho}_\mathrm{psq}\otimes\hat{\rho}_{\mathrm{vac},c})\hat{B}_\mathrm{HR}^\dagger$, where $\hat{\rho}_{\mathrm{vac},c}$ corresponds to the vacuum input at the idle port.
As the projector representing the APD signal is $\hat{\Pi}=\sum\limits_j \ket{1}_j\bra{1} _j$, postselection on the signal evolves the state into
\begin{eqnarray}\label{eq:density operator after postselction}
\begin{split}
\hat{\rho}_\mathrm{sub}=\frac{\text{tr}_{c}\big[\big(\mathds{1}\otimes\Pi\big) \, \hat{\rho}'\big]}{\text{tr}\big[\big(\mathds{1}\otimes\Pi\big) \, \hat{\rho}' \big]}=\frac{\sum\limits_j \hat{b_j} \hat{\rho}_\mathrm{psq} \hat{b}_j^\dagger}{\text{tr}\Big[\sum\limits_j \hat{b_j} \hat{\rho}_\mathrm{psq} \hat{b}_j^\dagger\Big]}=\frac{\Big(\sum\limits_{j=1}^4\bigotimes\limits_{k=1}^4 \hat{b}_j \hat{\rho}_k \hat{b}_j^\dagger\Big)\otimes \hat{\rho}_\mathrm{vac}}{\sum\limits_{j=1}^4 \sinh^2 r_j }\;,
\end{split}
\end{eqnarray}
where equation \eqref{eq:BMreduction} has been used, with $\hat{\rho}_\mathrm{vac}$ denoting the effective vacuum of high-order modes of the reflected light.

Probabilities to get $0$ and $1$, $P(0)$ and $P(1)$, from the outcome of photon number measurement for $\hat{A}^{(d)}(t_d)$ mode are shown in Fig.~\ref{Fig:Single photon subtraction}b to illustrate the effect of the postselection. For instance, the dynamics of the TRWF value at the origin depicted in Fig.~\ref{Fig:MPandNegativity}b can be understood from them.
Equation \eqref{eq:density operator after postselction} implies that the probability of the photon subtraction to occur in the mode $j$ is proportional to $\sinh^2 r_j$.
For the photon-subtracted state, $P_\mathrm{sub}(1)$ can be approximated then by $T_1(t_d)$
since the first mode has much larger squeezing parameter than the other modes.
Near $-14\,$fs, the postselection increases the probability to detect a photon to $0.4$ whereas it leads to a decrease in the TRWF value at the origin to $W(0,0;t_d)\approx \frac{1}{\pi}\left[P_\mathrm{sub}(0)-P_\mathrm{sub}(1)\right] \approx \frac{1}{\pi}\left[1-2 T_1(t_d)\right]$.

\section{Number of required measurements for the Gram-Charlier expansion}
$N$th-order statistical moments of any generalized quadrature can be expressed as linear combinations of $\langle \hat{X}^n(t_d) \hat{P}^m(t_d) \rangle_S$ where $n+m=N$, i.e. of the $N$th-order symmetrized moments of $\hat{X}(t_d)$ and $\hat{P}(t_d)$. Considering a vector space spanned by $\langle \hat{X}^n(t_d) \hat{P}^m(t_d) \rangle_S$, the $N$th-order statistical moment at phase $\phi$ represented in this basis gives
\begin{equation}\label{eq:vector_of_moments}
\left[\cos^N\phi, \tensor*[_N]{C}{_1} \cos^{N-1}\phi \sin\phi ,\ldots,\tensor*[_N]{C}{_k} \cos^{N-k}\phi \sin^k \phi,\ldots,\tensor*[_N]{C}{_{N-1}} \cos\phi \sin^{N-1} \phi, \sin^N \phi \right],
\end{equation}
where $\phi\in[0,\pi)$ and $\tensor*[_N]{C}{_k}=\begin{pmatrix} n \\ k\end{pmatrix}$ denote binomial coefficients. Linear independence of such vectors for $N+1$ different phases ensures that measurements of quadrature moments up to the $N$th order for $N+1$ phases are sufficient for approximating the TRWF by the $N$th-order Gram-Charlier expansion, i.e. taking into account all terms with $n+m\leq N$ in equation \eqref{eq:GCexpansionofW}.
Matrix $G_N$ constructed by combining these vectors of $N+1$ phases,
\begin{equation}
G_N=\left[ \begin{array}{ccccc}
\cos^N  \phi_1 & \cdots & \tensor*[_N]{C}{_k} \cos^{N-k}\phi_1 \sin^k \phi_1 & \cdots & \sin^N \phi_1 \\
\vdots & & \ddots & & \vdots\\
\cos^N  \phi_{N+1} & \cdots & \tensor*[_N]{C}{_k} \cos^{N-k}\phi_{N+1} \sin^k \phi_{N+1} & \cdots & \sin^N \phi_{N+1} \\
\end{array}\right],
\end{equation}
represents a linear transformation from the symmetrized quadrature moments to the generalized ($\phi$-dependent) quadrature moments: 
\begin{equation}
\left[ \begin{array}{c}\big\langle \hat{X}^N_{\phi_1}(t_d) \big\rangle \\ \vdots \\ \big\langle \hat{X}^N_{\phi_{N+1}}(t_d) \big\rangle \end{array} \right]
=G_N
\left[ \begin{array}{c}\big\langle \hat{X}^N(t_d) \big\rangle_S \\ \vdots \\  \big\langle \hat{P}^N(t_d) \big\rangle_S \end{array} \right].
\end{equation}
Proof of the linear independence of the vectors determined by equation \eqref{eq:vector_of_moments} can be done by showing $\det G_N \neq0$. This then would also mean the $N$th-order symmetrized moments can be obtained by applying $G_N^{-1}$ to $\left[ \big\langle \hat{X}^N_{\phi_1}(t_d) \big\rangle, \big\langle \hat{X}^N_{\phi_2}(t_d) \big\rangle,\ldots,\big\langle \hat{X}^N_{\phi_{N+1}}(t_d) \big\rangle \right]^\top$.

If we do not include $\phi=\frac{\pi}{2}$, the transformation matrix $G_N$ can be decomposed into a product of the Vandermonde matrix and diagonal matrices
\begin{equation}
\begin{split}
&G_N=
\\
&\mathrm{diag}(\cos^N\phi_1,\ldots,\cos^N\phi_k,\ldots,\cos^N\phi_{N+1})
\left[ \begin{array}{ccccc}
1 & \cdots & \tan^k \phi_1 & \cdots & \tan^N \phi_1 \\
\vdots & & \ddots & & \vdots\\
1 & \cdots & \tan^k \phi_{N+1} & \cdots & \tan^N \phi_{N+1} \\
\end{array} \right]\mathrm{diag}(1,\ldots,\tensor*[_N]{C}{_k},\ldots,1),
\end{split}
\end{equation}
where $\mathrm{diag}$ denotes the corresponding diagonal matrix.
From that we get
$\det G_N=\prod\limits_{j=1}^{N+1} \cos^N \phi_j \allowbreak \prod\limits_{1\leq i < j \leq N+1} (\tan\phi_j-\tan\phi_i)  \prod\limits_{j=0}^{N} \tensor*[_N]{C}{_k} \neq 0$ if all $\phi_i$ are different. If we include $\frac{\pi}{2}$ phase, $\phi_1=\frac{\pi}{2}$ without loss of generality, the decomposition becomes
\begin{equation}
\begin{split}
&G_N=
\\
&\mathrm{diag}(1, \cos^N\phi_2,\ldots,\cos^N\phi_k,\ldots,\cos^N\phi_{N+1})
\left[ \begin{array}{ccccc}
0 & \cdots & 0 & \cdots & 1\\
1 & \cdots & \tan^k \phi_2 & \cdots & \tan^N \phi_2 \\
\vdots & & \ddots & & \vdots\\
1 & \cdots & \tan^k \phi_{N+1} & \cdots & \tan^N \phi_{N+1} \\
\end{array} \right]\mathrm{diag}(1,\ldots,\tensor*[_N]{C}{_k},\ldots,1).
\end{split}
\end{equation}
This matrix also possesses the property $\det G_N=\prod\limits_{j=2}^{N+1} \cos^N \phi_j  (-1)^N \prod\limits_{2\leq i < j \leq N+1} (\tan\phi_j-\tan\phi_i)  \prod\limits_{j=0}^{N} \tensor*[_N]{C}{_k} \neq 0$ in case of all phases being different from each other. Therefore, the $N$th-order symmetrized moments can be obtained from measurements of  the $N$th-order quadrature moments for $N+1$ different phases.

Reconstruction of the TRWF based on the 2nd-order approximation of equation \eqref{eq:GCexpansionofW}, for instance, requires the first and second symmetrized moments. They can be extracted from measured quadrature moments via the inverse matrices
\begin{equation}
G^{-1}_1=\left[\begin{array}{cc}
-\frac{\sin\phi_2}{\sin\left(\phi_1-\phi_2\right)} & \frac{\sin\phi_1}{\sin\left(\phi_1-\phi_2\right)}
\\[0.2cm]
-\frac{\cos\phi_2}{\sin\left(\phi_2-\phi_1\right)} & \frac{\cos\phi_1}{\sin\left(\phi_2-\phi_1\right)}
\end{array}\right]
\end{equation}
and
\begin{equation}
G^{-1}_2=
\left[\begin{array}{ccc}
\frac{\sin\phi_2 \sin\phi_3}{\sin\left(\phi_1-\phi_2\right) \sin\left(\phi_1-\phi_3\right)} & \frac{\sin\phi_1 \sin\phi_3}{\sin\left(\phi_2-\phi_1\right) \sin\left(\phi_2-\phi_3\right)} & \frac{\sin\phi_1 \sin\phi_2}{\sin\left(\phi_3-\phi_1\right) \sin\left(\phi_3-\phi_2\right)}
\\[0.2cm] -\frac{\sin\left(\phi_2+\phi_3\right)}{2\sin\left(\phi_1-\phi_2\right) \sin\left(\phi_1-\phi_3\right)} & -\frac{\sin\left(\phi_1+\phi_3\right)}{2\sin\left(\phi_2-\phi_1\right) \sin\left(\phi_2-\phi_3\right)} & -\frac{\sin\left(\phi_1+\phi_2\right)}{2\sin\left(\phi_3-\phi_1\right) \sin\left(\phi_3-\phi_2\right)}
\\[0.2cm]
\frac{\cos\phi_2 \cos\phi_3}{\sin\left(\phi_1-\phi_2\right) \sin\left(\phi_1-\phi_3\right)} & \frac{\cos\phi_1 \cos\phi_3}{\sin\left(\phi_2-\phi_1\right) \sin\left(\phi_2-\phi_3\right)} & \frac{\cos\phi_1 \cos\phi_2}{\sin\left(\phi_3-\phi_1\right) \sin\left(\phi_3-\phi_2\right)}
\end{array}
\right].
\end{equation}
If we select $0, \frac{\pi}{2}$ and $\frac{\pi}{4}$ for the phases, the first symmetrized moments are found from
\begin{equation}
\left[\begin{array}{c}
\langle \hat{X}(t_d) \rangle_S
\\[0.1cm]
\langle \hat{P}(t_d) \rangle_S
\end{array}\right]
=\left[\begin{array}{cc}
 1 & 0
\\[0.1cm]
0 & 1
\end{array}\right]
\left[\begin{array}{c}
\langle \hat{X}(t_d) \rangle
\\[0.1cm]
\langle \hat{X}_{\frac{\pi}{2}}(t_d) \rangle
\end{array}\right],
\end{equation}
whereas the second symmetrized moments can be obtained from
\begin{equation}
\left[\begin{array}{c}
\langle \hat{X}^2(t_d) \rangle_S
\\[0.1cm]
\langle \hat{X}(t_d) \hat{P}(t_d) \rangle_S
\\[0.1cm]
\langle \hat{P}^2(t_d) \rangle_S
\end{array}\right]
=\left[\begin{array}{ccc}
 1 & 0 & 0
\\[0.2cm]
-\frac{1}{2} & -\frac{1}{2} & 1
\\[0.2cm]
0 & 1 & 0
\end{array}\right]
\left[\begin{array}{c}
\langle \hat{X}^2(t_d) \rangle
\\[0.1cm]
\langle \hat{X}^2_{\frac{\pi}{2}}(t_d) \rangle
\\[0.1cm]
\langle \hat{X}^2_{\frac{\pi}{4}}(t_d) \rangle
\end{array}\right].
\end{equation}
Such process can be elaborated as follows. Measurements of quadrature moments for $0$-phase give $\langle\hat{X}(t_d)\rangle$ and $\langle\hat{X}(t_d)^2\rangle$. Measurements of quadrature moments for $\frac{\pi}{2}$-phase give $\langle\hat{P}(t_d)\rangle$ and $\langle\hat{P}(t_d)^2\rangle$. The remaining cross moment can be extracted from the variance of the $\frac{\pi}{4}$-rotated quadrature: $\big\langle\frac{\hat{X}(t_d)\hat{P}(t_d)+\hat{P}(t_d)\hat{X}(t_d)}{2}\big\rangle=\big\langle \big(\frac{\hat{X}(t_d)+\hat{P}(t_d)}{\sqrt{2}}\big)^2\big\rangle-\frac{1}{2}\langle\hat{X}(t_d)^2\rangle-\frac{1}{2}\langle\hat{P}(t_d)^2\rangle$. In a similar way, $W(x,p;t_d)$ can be approximated by the $N$th order of equation \eqref{eq:GCexpansionofW}
from
measurements of quadrature moments up to the $N$th order for $N+1$ phases at each time. Although this method allows for reconstruction of the TRWF relying on an arbitrary choice of the phases, maintaining high precision of the measured moments when transforming to the symmetrized moments may require these phases to differ significantly from each other. Such a choice assures that 
the conditional number\cite{Chapra2015} of $G^{-1}_N$ stays relatively small.

After getting symmetrized moments from the general quadrature moments, the TRWF can be reconstructed from the Gram-Charlier expansion \eqref{eq:GCexpansionofW} with its coefficients given by symmetrized moments, see Table \ref{table:Gram-Charlier coefficients}.

\begin{table}[h!]
\centering
\begin{tabular}{||c|c|c|c|c|c|c||}
 \hline
 \multicolumn{6}{|c|}{$C_{nm}$} \\ [0.5ex]
 \hline\hline
 \diagbox[width=1cm]{n}{m} & 0 & 1 & 2 & 3 & 4\\[4pt] \hline\xrowht{20pt}
0 & 1 & $\langle \hat{P} \rangle_S$  & $\frac{\langle \hat{P}^2 \rangle_S}{2} -\frac{1}{4}$ & $\frac{\langle \hat{P}^3 \rangle_S}{6} -\frac{\langle \hat{P} \rangle_S}{4}$ & $\frac{\langle \hat{P}^4 \rangle_S}{24}-\frac{\langle \hat{P}^2 \rangle_S}{8} +\frac{1}{32}$ \\ \hline\xrowht{20pt}
1 & $\langle \hat{X} \rangle_S$ & $\langle \hat{X}\hat{P} \rangle_S$  & $\frac{\langle \hat{X}\hat{P}^2 \rangle_S}{2} -\frac{\langle \hat{X} \rangle_S}{4}$ & $\frac{\langle \hat{X}\hat{P}^3 \rangle_S}{6}  -\frac{\langle \hat{X}\hat{P} \rangle_S}{4} $ &  \\ \hline\xrowht{20pt}
2 & $\frac{\langle \hat{X}^2 \rangle_S}{2} -\frac{1}{4}$ & $\frac{\langle \hat{X}^2\hat{P} \rangle_S}{2} -\frac{\langle \hat{P} \rangle_S}{4}$ & $\frac{\langle \hat{X}^2\hat{P}^2 \rangle_S}{4}-\frac{\langle \hat{X}^2 \rangle_S }{8}-\frac{\langle \hat{P}^2 \rangle_S }{8}+\frac{1}{16}$ & & \\ \hline\xrowht{20pt}
3 & $\frac{\langle \hat{X}^3 \rangle_S}{6} -\frac{\langle \hat{X} \rangle_S}{4}$ & $\frac{\langle \hat{X}^3\hat{P} \rangle_S}{6}-\frac{\langle \hat{X}\hat{P} \rangle_S}{4}$ & & & \\ \hline\xrowht{20pt}
4 & $\frac{\langle \hat{X}^4 \rangle_S}{24} -\frac{\langle \hat{X}^2 \rangle_S}{8} +\frac{1}{32}$ & & & & \\ \hline
\end{tabular}
\caption{Coefficients of the Gram-Charlier expansion in terms of the symmetrically ordered moments of $\hat{X}(t_d)$ and $\hat{X}(t_d)$ up to $n+m=4$th order}
\label{table:Gram-Charlier coefficients}
\end{table}

\section{Precision of the reconstruction and higher orders of the Gram-Charlier expansion}

\begin{figure}[h]
    \includegraphics[width=18cm]{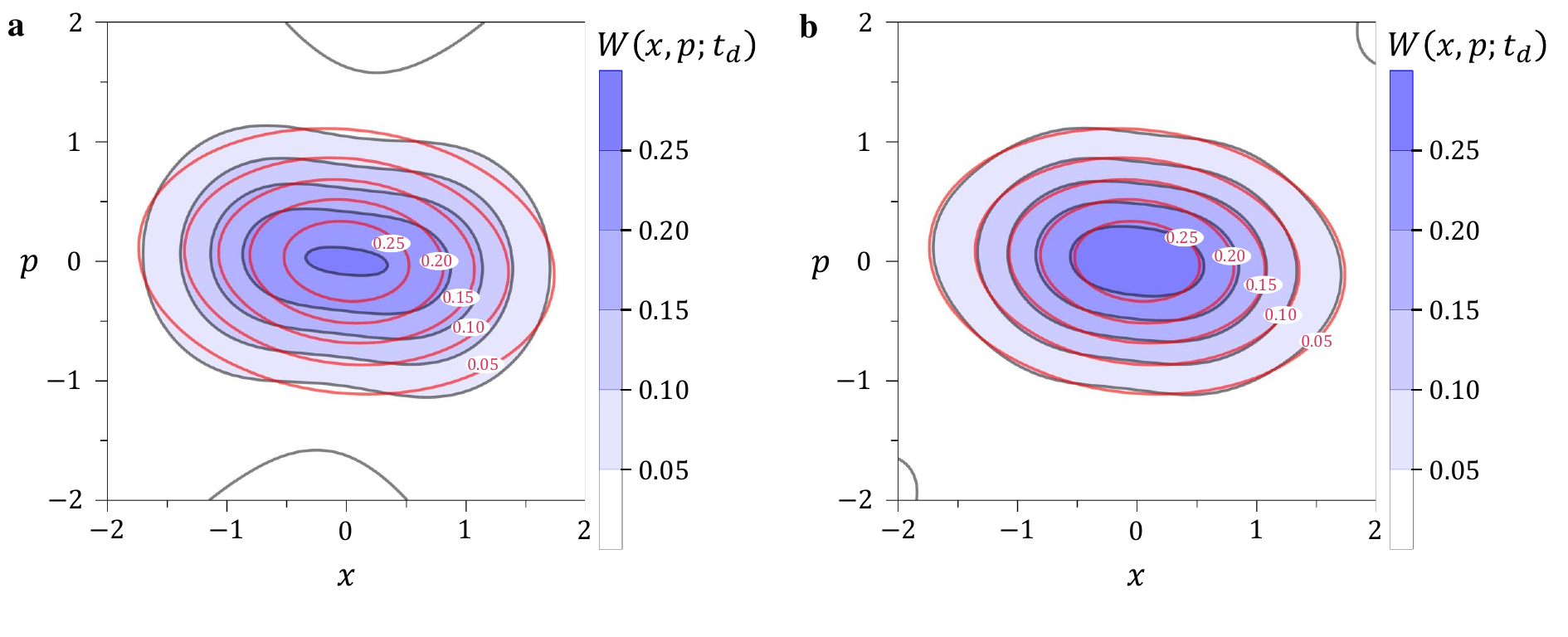}
    \centering
    \caption{\label{Fig:GC expansion when r 20}
    \textbf{Reconstruction of the TRWF from the Gram-Charlier expansion for high squeezing.} Colour schemes of the exactly calculated TRWF and TRWF reconstructed from the Gram-Charlier expansion are the same as in
    Fig.~\ref{Fig:GCexpansion}. The distributions are depicted at the time moment $t_d=-22.6\,$fs when $D_\mathrm{HS}\left(W(x,p;t_d),W_\mathrm{vac}(x,p)\right)$ becomes maximum. Reconstruction from the moments up to the 2nd order (\textbf{a}) and 6th order (\textbf{b}). Difference between the exact and reconstructed TRWF is quantified here by
    $D_\mathrm{HS}=0.0103$ ($D_\mathrm{HS}=0.0011$) for \textbf{a} (\textbf{b}). Parameters: $r_\mathrm{eff}=20$, $\delta_d=16\,$fs, and $\delta_p=8\,$fs (cf. Sections \ref{Methods:Generation} and \ref{Methods:Sampling} of Methods).}
\end{figure}		

The 2nd-order Gram-Charlier expansion is employed to reconstruct the TRWF of ultrafast squeezed state and single-photon-subtracted state produced from it in Fig.~\ref{Fig:GCexpansion}.
Within this level approximation the procedure works well for moderate squeezing, indicated by minute values of the distance between the original and reconstructed distributions $D_\mathrm{HS}$.
However, for the case of high squeezing, which is anyway problematic in terms of the generation \cite{KizmannSubcycle},  higher orders of the expansion are required.
For the reason of illustration, we tested the case of $r_\mathrm{eff}=20$.
%
We calculated the TRWF of such squeezed state analytically, using the probe pulse duration $\delta_p=8\,$fs and the shape of 
the driving field as in
Fig.~\ref{Fig:GCexpansion}, and then checked how its reconstruction would look like based on the 2nd-order (Fig.~\ref{Fig:GC expansion when r 20}a) and then 6th-order (Fig.~\ref{Fig:GC expansion when r 20}b) Gram-Charlier expansion. Figure \ref{Fig:GC expansion when r 20} shows a comparison
for
a single time moment selected so that $D_\mathrm{HS}$ between the TRWF and the vacuum distribution $W_\mathrm{vac}(x,p)$ becomes maximum.
Although the TRWF reconstructed based on the 2nd-order expansion in Fig.~\ref{Fig:GC expansion when r 20}a still well reflects the degree of squeezing and angle of its axis, it exhibits concave parts which cannot be found in the exact TRWF. The deviations become suppressed
when reconstructing the TRWF from higher moments, as in Fig.~\ref{Fig:GC expansion when r 20}b. In the case of  inversion symmetric $W(x,p;t_d)$ like normal distributions, $C_{nm}$ in equation \eqref{eq:GCexpansionofW} vanish when $n+m$ is odd. Therefore,
corrections appear only in even orders of the expansion. Figure ~\ref{Fig:GC expansion when r 20}b illustrates the result of the 6th order.


\section{Ultrabroadband single photon}

\begin{figure}[h]\includegraphics[width=18cm]{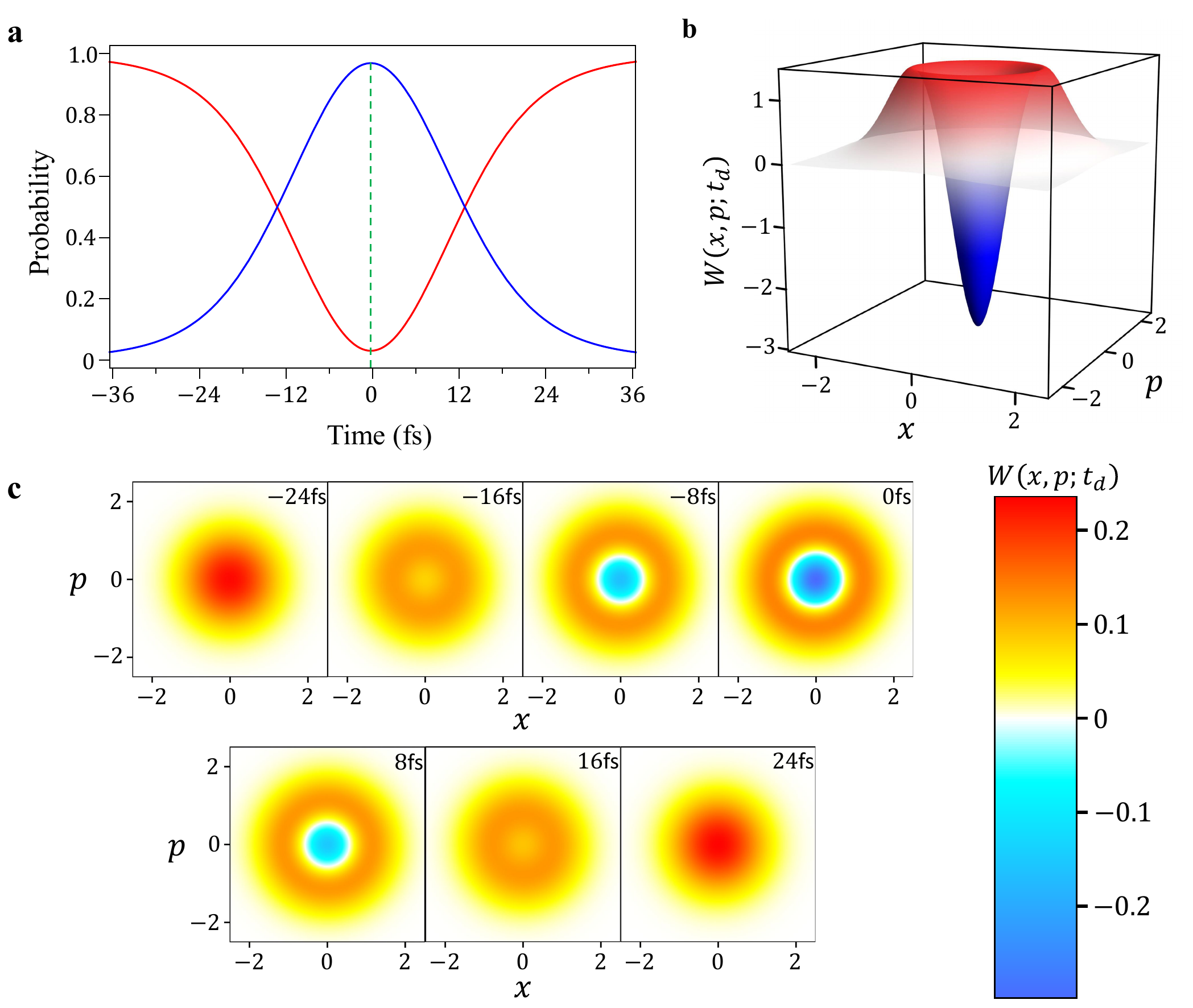}
    \centering
    \caption{\label{Fig:Ultrafast single-photon state}
    \textbf{Subcycle dynamics of a photon-subtracted state for weak squeezing.} \textbf{a,} Probabilities of the photon number measurement corresponding to the vacuum outcome (red) and  single-photon outcome (blue). \textbf{b,} TRWF at the time moment when the probability of the single-photon outcome reaches its maximum (i.e. green dotted line in \textbf{a}: $t_d=-0.26\,$fs where the probability to detect a single photon amounts to $0.969$). \textbf{c,} Snapshots of the dynamics of the TRWF for the ultrafast single photon. Parameters: $r_\mathrm{eff}=0.1$, $\delta_d=16\,$fs, and $\delta_p=18.47\,$fs (cf. Sections \ref{Methods:Generation} and \ref{Methods:Sampling} of Methods).}
\end{figure}

Similar to the case of a single mode, an ultrabroadband single photon can be generated upon a photon subtraction from a weak squeezed state. For that, one can implement the scheme of Fig.~\ref{Fig:Subcycle Tomography by BHD}a with a small amplitude of the driving field.
A weak ultrafast squeezing with $r_\mathrm{eff}=0.1$ is almost equivalent to the generation of a two-photon state, since the squeezing parameters of its first two principal modes are $r_1=0.00961$ and $r_2=0.00035$ (so that the weakly squeezed first mode dominates). After the photon subtraction, the remaining single photon is distributed in time centered at $t_d=-0.26$\,fs,
as can be seen in Fig.~\ref{Fig:Ultrafast single-photon state}a. The phase-space properties of this photon
are well captured by the TRWF, shown in Fig.~\ref{Fig:Ultrafast single-photon state}b for $t_d=-0.26$\,fs.
The dynamics of the TRWF in Fig.~\ref{Fig:Ultrafast single-photon state}c reveals the appearance and disappearance of the transient single photon, in this case quite symmetrically in time.

\section{Extraction of the dominant field mode from the TRWF}\label{sup:field mode reconstruction}

\begin{figure}[h]\includegraphics[width=18cm]{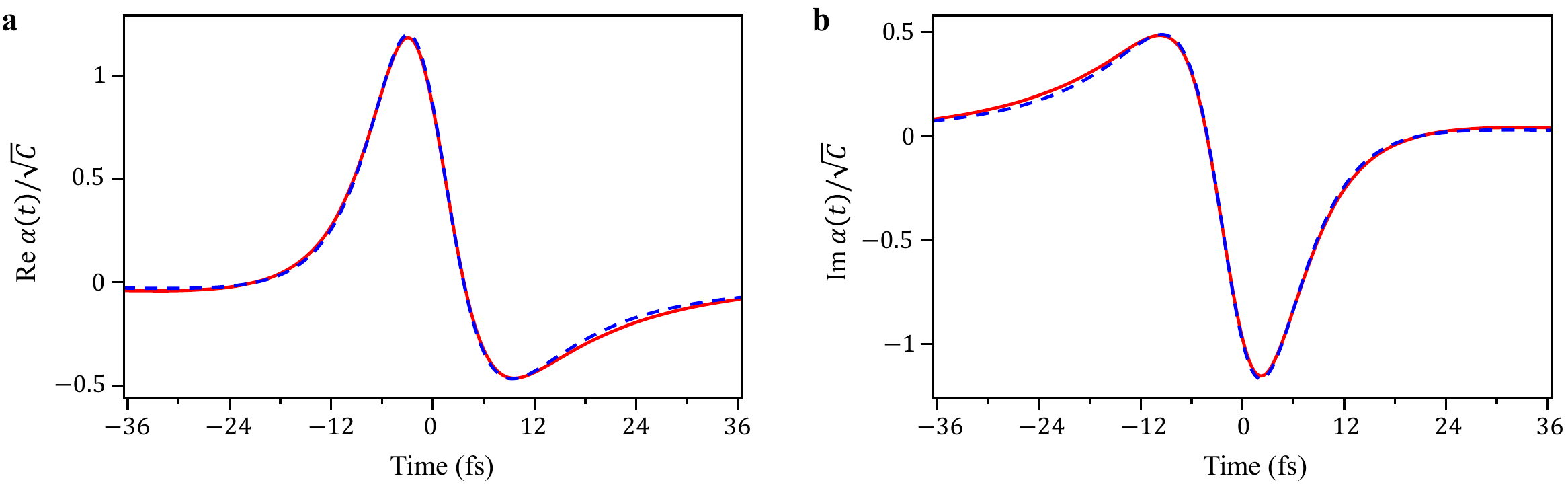}
    \centering
    \caption{\label{Fig:Field mode reconstruction}
    \textbf{Reconstruction of the most dominant field mode of an ultrafast weakly squeezed state
    based on the subcycle tomography.} Real (\textbf{a}) and imaginary (\textbf{b}) parts of the reconstructed (red solid line) and exact (blue dashed line) dominant field mode. The constant $C$ used for the normalization is introduced in Section \ref{Methods:Sampling} of Methods. Parameters: $r_\mathrm{eff}=0.1$, $\delta_d=16\,$fs, and $\delta_p=5.8\,$fs (cf. Sections \ref{Methods:Generation} and \ref{Methods:Sampling} of Methods).}
\end{figure}

Observed quantum statistics from the subcycle tomography are determined by the dominant field modes and their corresponding states. Therefore, it may be possible to deduce the field modes and their quantum states
from the TRWF. As a simple clarifying example, we consider weakly squeezed ultrabroadband states with $r_\mathrm{eff}=0.1$ and demonstrate how to extract them from the measured TRWF. As a consequence of negligible squeezing parameters for all principal modes except for the first one, $r_1=0.00961$ ($r_2=0.00035$ is already much smaller), the observed covariance matrix can be expressed as
\begin{eqnarray}\label{eq:Single-Mode Covariancematrix}
\boldsymbol{\Sigma}(t_d)=\frac{1}{2}\left[\theta^2_\mathrm{vac}(t_d)\,\mathbf{I} + \mathbf{O}(t_d) \,\mathbf{S} \,\mathbf{O}^\top(t_d)\right],
\end{eqnarray}
where $\mathbf{I}=\left[\begin{array}{cc}1 & 0 \\ 0 & 1\end{array}\right]$, $\mathbf{S}=\left[\begin{array}{cc}e^{2r} & 0 \\ 0 & e^{-2r}\end{array}\right]$, $\mathbf{O}_j(t_d)=\left[\begin{array}{cc}\Re\theta(t_d) & -\Im\theta(t_d)\\ \Im\theta(t_d) & \Re\theta(t_d)\end{array}\right]$, and $\theta_\mathrm{vac}(t_d)=\sqrt{1-\lvert \theta(t_d) \rvert^2}$ [cf. equation \eqref{eq:Covariancematrix}]. Then $\theta(t_d)$ and $r$ are determined by the eigenvalues $V_\mathrm{max}$, $V_\mathrm{min}$ of $\boldsymbol{\Sigma}(t_d)$ and the polar angle $\phi$ of the eigenvector corresponding to $V_\mathrm{max}$:
\begin{eqnarray}\label{eq:Single-Mode theta and r}
\theta (t_d)=\sqrt{\frac{ ( 2 V_\mathrm{max}-1) (1-2 V_\mathrm{min})}{4 (V_\mathrm{max}+V_\mathrm{min}-1)}}e^{i\phi},\qquad r=\frac{1}{2}\ln \left( \frac{2 V_\mathrm{max}-1}{1-2 V_\mathrm{min}} \right).
\end{eqnarray}
Based on the first equation of \eqref{eq:Single-Mode theta and r}, we reconstructed the field mode $\alpha (t)$ by the deconvolution of $\theta(t_d)$ with the gating function, as shown in Fig.~\ref{Fig:Field mode reconstruction}. The squeezing parameter $r$ evaluated based on the second of equations \eqref{eq:Single-Mode theta and r} amounts to $0.00955$. One can see that in spite of a considerable probe pulse duration $t_p=5.8\,$fs, with respect to the optical cycle, our subcycle tomography is capable to deduce the field mode and its squeezing parameter quite accurately.


\providecommand{\noopsort}[1]{}\providecommand{\singleletter}[1]{#1}%

\end{document}